\documentclass[twocolumn,showpacs,preprintnumbers,amsmath,amssymb, superscriptaddress, showkeys]{revtex4}

\usepackage{graphicx} 
\usepackage{dcolumn} 
\usepackage{bm} 
\usepackage{color} 
\usepackage{slashed}

\newcommand{\mrm}[1]{\mathrm{#1}}

\begin{document}

\preprint{published in Phys.\ Rev.\ D}

\title{Cosmological implications of a\\ Dark Matter self-interaction energy density}

\author{Rainer Stiele}
 \email{R.Stiele@ThPhys.Uni-Heidelberg.DE}
\affiliation{Institute for Theoretical Physics, Heidelberg University, Philosophenweg 16, D-69120 Heidelberg, Germany}
\affiliation{Institute for Theoretical Physics, Goethe University Frankfurt, Max-von-Laue-Stra{\ss}e 1, D-60438 Frankfurt am Main, Germany}

\author{Tillmann Boeckel}
 \email{T.Boeckel@ThPhys.Uni-Heidelberg.DE}
\author{J\"urgen Schaffner-Bielich}
 \email{Schaffner-Bielich@Uni-Heidelberg.DE}
\affiliation{Institute for Theoretical Physics, Heidelberg University, Philosophenweg 16, D-69120 Heidelberg, Germany}

\date{\today}

\begin{abstract}
We investigate cosmological constraints on an energy density contribution of elastic dark matter self-interactions characterized by the mass of the exchange particle $m_\mrm{SI}$ and coupling constant $\alpha_\mrm{SI}$. Because of the expansion behaviour in a Robertson-Walker metric we investigate self-interacting dark matter that is warm in the case of thermal relics. The scaling behaviour of dark matter self-interaction energy density ($\varrho_\mrm{SI}\propto{a^{-6}}$) shows that it can be the dominant contribution (only) in the very early universe. Thus its impact on primordial nucleosynthesis is used to restrict the interaction strength $m_\mrm{SI}/\!\sqrt{\alpha_\mrm{SI}}$, which we find to be at least as strong as the strong interaction. Furthermore we explore dark matter decoupling in a self-interaction dominated universe, which is done for the self-interacting warm dark matter as well as for collisionless cold dark matter in a two component scenario. We find that strong dark matter self-interactions do not contradict super-weak inelastic interactions between self-interacting dark matter and baryonic matter ($\sigma_\mrm{A}^\mrm{SIDM}\ll\sigma_\mrm{weak}$) and that the natural scale of collisionless cold dark matter decoupling exceeds the weak scale ($\sigma_\mrm{A}^\mrm{CDM}>\sigma_\mrm{weak}$) and depends linearly on the particle mass. Finally structure formation analysis reveals a linear growing solution during self-interaction domination ($\delta\propto{a}$); however, only non-cosmological scales are enhanced.
\end{abstract}


\maketitle

\section{\label{sec:introduction}introduction\protect}
In the past decades high-precision observations allowed the development of a standard model of cosmology: $\Lambda$CDM. Its main statements are that we are living in a flat universe ($\Omega_\mrm{tot}^0={1.0052}\pm{0.0064}$), dominated by the {\textquoteleft}dark' components: dark energy ($\Omega_\mrm{DE}^0={0.721}\pm{0.015}$) and non-baryonic dark matter ($\Omega_\mrm{DM}^0={0.233}\pm{0.013}$) \cite{Hinshaw_2009}.\\
The necessity of a dark energy component comes from the acceleration of the universe expansion, inferred from a high redshift Hubble diagram of type Ia supernovae as standard candles \cite{Kowalski_2008} and radio galaxies as standard yardsticks \cite{Daly_2009}. A recent and impressive proof for the existence of dark matter (DM) can be deduced from observations of colliding galaxy clusters. Optical and near infrared observations of the galaxies, X-ray emission of the upheated intergalactic plasma and gravitational lensing of the mass distribution show the necessity of a non-visible DM component, which dominates the mass budget \cite{Clowe_2006, Bradac_2008}. An overview of DM physics and particle candidates can e.g.\ be found in recent reviews \cite{Baltz_2004, Bergstroem_2009_A, Bertone_2005, Taoso_2008}.

Numerical structure formation simulations in the $\Lambda$CDM framework (from one of the first and most popular \cite{Navarro_1996} to the most recent \cite{Springel_2008, Navarro_2010}) show an impressive agreement with observations and have therefore become a cornerstone of modern cosmology. Nevertheless they also reveal two shortcomings that are worth being taken seriously. First, simulations predict scale-independently a large number of substructures in collisionless cold dark matter  (CDM) halos, which exceed on galactic scales clearly the number of yet observed Milky Way satellites \cite{Klypin_1999, Moore_1999}. One explanation is that reionization could prevent formation of visible baryonic structures in the smallest DM halos (e.g.\ \cite{Shapiro_2004}). Second, simulations show cusps in the center of collisionless CDM halo density profiles. But observations of dwarf spheroidal galaxies -- which have a huge mass-to-light ratio and hence are objects suited to study DM properties without perturbing baryonic effects, rotation curves of high spatial resolution and large extension of low luminosity spiral galaxies, and the universal rotation curve for spiral galaxies indicate a constant DM halo core density (e.g.\ \cite{Gilmore_2007, Gentile_2004, Salucci_2007}). An overview about processes that might lead from intrinsic cuspy CDM distributions to the observed cored ones gives e.g.\ Ref.\ \cite{deBlok_2010}.

An idea to avoid both mismatches of the CDM scenario is to introduce strong elastic DM self-interactions \cite{Spergel_2000}. A recent overview concerning collisional DM is given in Ref.\ \cite{Taoso_2008}. Here, we want to concentrate on the most important facts that are also relevant for this work. The original self-interaction strength proposed by Ref.\ \cite{Spergel_2000} is $\sigma_\mrm{SI}/m_\mrm{DM}=0.45-450\,\mrm{cm^2/g}$ (self-interaction cross-section over DM particle mass). But in Ref.\ \cite{Yoshida_2000} it was demonstrated that cross-sections generating reasonable dwarf galaxy cores predict too large galaxy cluster cores. Ref.\ \cite{Donghia_2003} showed that independent of the dependence on the halo velocity dispersion self-interacting cross-sections cannot solve the satellite problem accurately. The most reliable constraint on the self-interaction strength comes again from observations of colliding galaxy clusters \cite{Markevitch_2004, Randall_2008}. The nonexistence of an offset between the galaxy distribution and the gravitational lens mass peak, and the subcluster mass-to-light ratio allow to constrain $\sigma_\mrm{SI}/m_\mrm{DM}<0.7\,\mrm{cm^2/g}$. This result nearly completely rules out the formerly proposed self-interaction strength. The strongest limit on the collisional character of DM ($\sigma_\mrm{SI}/m_\mrm{DM}<0.02\,\mrm{cm^2/g}$) can be inferred from the observed ellipticity of DM halos and the property of DM collisions to make halos spherical \cite{Miralda_2002}, but one has to take into account its model dependence \cite{Markevitch_2004, Randall_2008}.

Another approach that avoids the satellite and cuspy halo problems is to provide the DM particles with a finite thermal streaming velocity, to achieve a cut-off in the power spectrum and smearing of the innermost, highest density halo regions \cite{Bode_2001, Sommer_2001} (see also \cite{Gilmore_2007}). This means to introduce warm dark matter (WDM) particles. A lower bound on the DM particle mass can be determined from the Lyman-$\alpha$ forest ($m_\mrm{WDM}\gtrsim4\,\mrm{keV}$, \cite{Viel_2008}) and gravitational lensing  ($m_\mrm{WDM}\gtrsim2.2\,\mrm{keV}$, \cite{Miranda_2007}) of high redshift quasars (given limits are for thermal relics). Ref.\ \cite{Boyarsky_2009} showed that these boundaries can be lowered considerably in a set-up of mixed cold and warm DM, which we also consider in this work. So we follow Ref.\ \cite{Boyanovsky_2008} in using $1-10\,\mrm{keV}$ as a typical mass range for WDM particles in the following.

Other solutions proposed are stronger CDM annihilations (\cite{Kaplinghat_2000}, but see also \cite{Beacom_2007}) or a coupling between quintessence dark energy and CDM (e.g.\ \cite{Baldi_2010}).

Present attempts to enlarge the phenomenology of DM physics are e.g.\ strong DM baryonic matter interactions \cite{Cyburt_2002} or a dark radiation (electromagnetism) between DM particles \cite{Ackerman_2009}.

We introduce in this work an energy density contribution of elastic dark matter self-interactions. Despite the fact that self-interacting dark matter (SIDM) may not solve the shortcomings of the collisionless approach, the motivation for this work is to explore new, interesting cosmological consequences of an additional energy density contribution of DM self-interactions within the above mentioned constraints. Interestingly enough, an interaction strength of $\sigma_\mrm{SI}/m\sim1\,\mrm{cm^2/g}$ still corresponds to strong interactions ($\sigma_\mrm{strong}\sim10\,\mrm{fm^2}$) between nucleon-like particles ($m\sim1\,\mrm{GeV}$).\\
In Sec.\ \ref{sec:SIDM model} we introduce our idea of a self-interaction energy density contribution $\varrho_\mrm{SI}$. Energy density scaling according to the Friedmann equations and its equation of state ($p_\mrm{SI}=\varrho_\mrm{SI}$) imply that the self-interaction contribution shows the steepest decrease with the scale factor ($\varrho_\mrm{SI}\propto{a^{-6}}$) and thus can (only) have a direct impact on the very early universe. Its proportionality to the SIDM particle density ($\varrho_\mrm{SI}\propto{n_\mrm{SIDM}^2}$) leads us to consider warm self-interacting dark matter (WSIDM) in the case of thermal relics to have the correct scaling behaviour ($n_\mrm{SIDM}\propto{a^{-3}}$). But this does not rule out a second collisionless CDM component.\\
After finally defining our parameter set, we use in Sec.\ \ref{sec:constraints} today's DM energy density $\Omega_\mrm{DM}^0$ to constrain the parameters characterising the SIDM particle properties and primordial nucleosynthesis limits on an additional energy density contribution to constrain the self-interaction strength $m_\mrm{SI}/\!\sqrt{\alpha_\mrm{SI}}$. We find that it depends inversely on the SIDM particle mass ($m_\mrm{SI}/\!\sqrt{\alpha_\mrm{SI}}\propto{1/m_\mrm{WDM}}$) but can be at least as strong as for strong interactions ($m_\mrm{SI}/\!\sqrt{\alpha_\mrm{SI}}\sim100\,\mrm{MeV}$).\\
In Sec.\ \ref{sec:decoupling} we analyse the consequences on DM decoupling in a universe dominated by the self-interaction energy density contribution. The annihilation cross-section of WSIDM $\sigma_\mrm{A}^\mrm{WDM}$ is inverse proportional to the elastic self-interaction strength ($\sigma_\mrm{A}^\mrm{WDM}\propto\sqrt{\alpha_\mrm{SI}}/m_\mrm{SI}$) and rather low ($\sigma_\mrm{A}^\mrm{WDM}\ll\sigma_\mrm{weak}$) while the natural scale for the annihilation cross-section of a collisionless CDM component $\sigma_\mrm{A}^\mrm{CDM}$ exceeds the weak scale and depends beside the self-interaction strength also on the particle mass $m_\mrm{CDM}$. This casts new light on the {\textquoteleft}WIMP miracle' and coincides with Fermi-LAT and PAMELA data (e.g.\ \cite{Grasso_2009, Bergstroem_2009_B}). We use the unitary bound and neutrino induced constraints on the DM annihilation cross-section to again limit the self-interaction strength.\\
Another consequence of an early self-interaction dominated epoch may concern structure formation. We show in Sec.\ \ref{sec:struct formation} that a relativistic analysis of linear perturbation theory reveals a linear growing solution $\delta\propto{a}$ of self-interaction dominated SIWDM and also of collisionless CDM in a mixed model during self-interaction domination. However, only non-cosmological scales ($M\lesssim10^{-3}M_\odot$) can be enhanced and a small observable effect could only be present with fine-tuned parameters.\\
Finally we summarize our results in Sec.\ \ref{sec:conclusions}.

\section{\label{sec:SIDM model}Self-Interaction energy density}
In the following, we analyse constraints and consequences of an energy density contribution from dark matter self-interactions $\varrho_\mrm{SI}$.\\
We describe two particle interactions between scalar bosons or fermions by the exchange of vector mesons. For a scalar field $\phi$ (fermionic field $\psi$) and a vector field $V_\mu$ the Lagrangian reads
\begin{subequations}
\begin{eqnarray}
 {\cal L}_\mrm{b}&=&{\cal D}_\mu^* \phi^* {\cal D}^\mu \phi - m_\mrm{b}^2 \phi^*\phi- \tfrac{1}{4} V_{\mu\nu} V^{\mu\nu} + \tfrac{1}{2}m_\mrm{v}^2 V_\mu V^\mu,\qquad\\
 {\cal L}_\mrm{f}&=&\bar{\psi}\left(i\slashed{\cal D}-m_\mrm{f}\right)\psi- \tfrac{1}{4} V_{\mu\nu} V^{\mu\nu} + \tfrac{1}{2}m_\mrm{v}^2 V_\mu V^\mu\;,
\end{eqnarray}
\end{subequations}
with $V_{\mu\nu}=\partial_\mu V_\nu -\partial_\nu V_\mu$. The boson (fermion) field is coupled to the vector field by a minimal coupling scheme
\begin{equation}
{\cal D}_\mu = \partial_\mu + i g_{\mrm{v} \phi(\psi)} V_\mu\;,
\end{equation}
where $g_{\mrm{v}\phi(\psi)}$ is the $\phi(\psi)$-$V$ coupling strength. We treat the vector field as a classical field. In homogeneous and isotropic matter the spatial components of the vector field vanish and the equation of motion for the scalar (fermion) field reads:
\begin{subequations}
\begin{eqnarray}
 \left[{\cal D}_{\mu} {\cal D}^\mu + m_\mrm{b}^2 \right] \phi(x)&=&0\\
 \left[i\slashed{\cal D}-m_\mrm{f}\right]\psi(x)&=&0
\end{eqnarray}
\end{subequations}
In the mean-field approximation, after expanding into plane waves, we obtain the dispersion relation:
\begin{equation}
 \omega_{\phi(\psi)} = \sqrt{\vec{k}^2+m_\mrm{b(f)}^2} + g_{\mrm{v} \phi(\psi)} V_0 
\end{equation}
Note that the vector interaction between the scalar (fermionic) particles is repulsive which ensures the overall stability of selfinteracting boson matter and avoids an enhancement of the annihilation cross-section due to the formation of bound states.\\
The number density of bosons (fermions)
\begin{subequations}
\begin{eqnarray}
 n_\mrm{b} = J_0 &=& i \phi^* \partial_0 \phi - i \left(\partial_0 \phi^*\right) \phi -2g_{\mrm{v}\phi}V_0\,\phi^*\phi\;,\\
 n_\mrm{f} = J_0 &=& \bar{\psi} \gamma_0 \psi\;,
\end{eqnarray}
\end{subequations}
is just the source term for the vector field that is determined from the equation:
\begin{subequations}
\label{eq:eomV}
\begin{eqnarray}
 m_\mrm{v}^2 V_0 &=&  g_{\mrm{v}\phi} \left( i \phi^* \partial_0 \phi - i \left(\partial_0 \phi^*\right) \phi -2g_{\mrm{v}\phi}V_0\,\phi^*\phi\right)\nonumber\\
  &&=  g_{\mrm{v}\phi}\,n_\mrm{b}\\
 m_\mrm{v}^2 V_0 &=&  g_{\mrm{v}\psi} \bar{\psi} \gamma_0 \psi = g_{\mrm{v}\psi}\,n_\mrm{f}
\end{eqnarray}
\end{subequations}
The total energy density of the boson (fermion) matter can be determined from the energy-momentum tensor
\begin{subequations}
\begin{eqnarray}
 \varrho_\mrm{b} &=& \varrho\mrm{_b^{free}}  + \frac{1}{2}m_\mrm{v}^2 V_0^2
 = \varrho\mrm{_b^{free}}  + \frac{g_{\mrm{v}\phi}^2}{2 m_\mrm{v}^2}\,n_\mrm{b}^2\\
 \varrho_\mrm{f} &=& \bar{\psi} \gamma^0 \left( i \partial_0-g_{\mrm{v}\psi} V_0\right)\psi + \frac{1}{2}m_\mrm{v}^2 V_0^2\nonumber\\
 && =  \bar{\psi} \gamma^0 \left( i \partial_0-g_{\mrm{v}\psi} V_0\right)\psi + \frac{g_{\mrm{v}\psi}^2}{2 m_\mrm{v}^2} \,n_\mrm{f}^2\;,
\end{eqnarray}
\end{subequations}
where the equation of motion for the vector field (\ref{eq:eomV}) has been used. The energy density of free boson matter obeys for the lowest energy mode $k=0$, $\varrho\mrm{_b^{free}}= 2 m_\mrm{b}^2\,\phi^*\phi = m_\mrm{b}\,n_\mrm{b}$.\\
The pressure is given by:
\begin{subequations}
\begin{eqnarray}
 p_\mrm{b} &=& p\mrm{_b^{free}} + \frac{1}{2}m_v^2 V_0^2 = p\mrm{_b^{free}} + \frac{g_{\mrm{v}\phi}^2}{2 m_\mrm{v}^2} \,n_\mrm{b}^2\\
 p_\mrm{f} &=& \frac{1}{3}\bar{\psi}\left[\gamma^0 \left( i \partial_0-g_{\mrm{v}\psi} V_0\right)-m_\mrm{f}\right]\psi+ \frac{1}{2}m_\mrm{v}^2 V_0^2\nonumber\\
 &&=  \frac{1}{3}\bar{\psi}\left[\gamma^0 \left( i \partial_0-g_{\mrm{v}\psi} V_0\right)-m_\mrm{f}\right]\psi+\frac{g_{\mrm{v}\psi}^2}{2 m_\mrm{v}^2}\,n_\mrm{f}^2\quad
\end{eqnarray}
\end{subequations}
The pressure of free boson (fermion) matter fulfills $p^\mrm{free}=\varrho^\mrm{free}/3$ while the particles are relativistic, and for the lowest energy mode $k=0$ the total pressure is just given by the vector field contribution.\\
The form of the interaction is equal to the one used for investigating implications of interactions of fermions and bosons on compact stars in Refs. \cite{Narain_2006, Agnihotri_2009}.

For simplicity we denote the particle masses $m_\mrm{SIDM}\equiv{m_\mrm{b(f)}}$, $m_\mrm{SI}\equiv{m_\mrm{v}}$, and define the coupling constant $\alpha_\mrm{SI}\equiv{g_\mrm{v\phi(\psi)}^2/2}$, so that the energy density contribution from dark matter self-interactions reads
\begin{equation}
 \label{eq:rhoSI}
 \varrho_\mrm{SI}=\frac{\alpha_\mrm{SI}}{m_\mrm{SI}^2}\,n_\mrm{SIDM}^2=p_\mrm{SI}\;,
\end{equation}
with $m_\mrm{SI}/\!\sqrt{\alpha_\mrm{SI}}$ as the energy scale of the self-interaction.
This scale can also be interpreted as the vacuum expectation value of the Higgs field of the interaction. For weak interactions the interaction strength is given by Fermi's constant $1/\!\sqrt{G_\mrm{F}}$ or the vacuum expectation of the Higgs field generating the masses of the W and Z bosons, so that $m_\mrm{weak}/\!\sqrt{\alpha_\mrm{weak}}\sim300\,\mrm{GeV}$. Correspondingly, the strength of low energy strong interactions, quantum chromodynamics (QCD), is controlled by the pion decay constant as $1/f_\pi^2$ in chiral perturbation theory, with $f_\pi$ being the vacuum expectation value of the sigma field. Hence, for strong interactions the typical interaction energy scale is $m_\mrm{strong}/\!\sqrt{\alpha_\mrm{strong}}\sim100\,\mrm{MeV}$ \cite{Narain_2006}.\\
Note that according to the underlying propagator Eq.\ (\ref{eq:rhoSI}) is valid only when $m_\mrm{SI}>5\,T_\mrm{SIDM}$, so when the vector meson is non-relativistic. Otherwise the self-interaction energy density contribution scales like radiation ($\varrho_\mrm{SI}\propto{n_\mrm{SIDM}^2/T_\mrm{SIDM}^2}\propto{T_\mrm{SIDM}^4}$). Once $m_\mrm{SI}/\!\sqrt{\alpha_\mrm{SI}}$ is fixed, this can also be used as a boundary condition on the coupling constant for given $T_\mrm{SIDM}$. A corresponding discussion is given in Appendix \ref{asec:alphaSI}. Please also note that the coupling constant $\alpha_\mrm{SI}$ defined here differs from the common definition by a factor $2\pi$, e.g.\ $\alpha_\mrm{s}=g_\mrm{s}^2/(4\pi)$.\\
The equation of state (\ref{eq:rhoSI}) of the self-interaction $p_\mrm{SI}=\varrho_\mrm{SI}$ represents the stiffest possible equation of state consistent with causality.

\begin{figure}[h]
 \centering
 \includegraphics[width=0.47\textwidth]{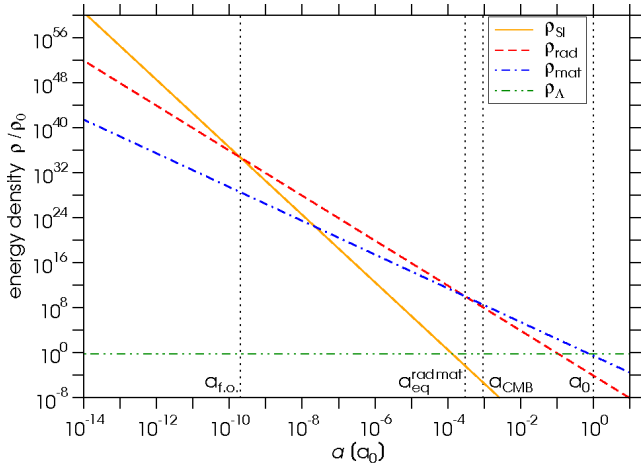}
 \caption[]{Double-logarithmic plot of the evolution of different energy density contributions with the scale factor $a$. $a_\mrm{CMB}$ is the scale factor at photon decoupling, $a_\mrm{eq}^\mrm{rad\,mat}$ at radiation-matter equality and $a_\mrm{f.o.}$ at the freeze-out of the neutron to proton number ratio. $\varrho_\mrm{SI}$ is fixed so that self-interaction--radiation equality is at freeze-out of the neutron to proton number ratio.}
 \label{fig:rhos_scaling}
\end{figure}

The expansion behaviour of the universe in a Robertson-Walker metric is described by the Friedmann equations
\begin{equation}
 \frac{\mrm{d}\varrho}{\mrm{d}a}=-3\,\frac{\varrho+p}{a}\;,
\end{equation}
where $a$ is the scale factor, implying that the self-interaction energy density contribution shows the steepest decrease (Fig.\ \ref{fig:rhos_scaling}):
\begin{equation}
 \varrho_\mrm{SI}\propto{a^{-6}}
\end{equation}
So, the universe could be in a self-interaction dominated epoch prior to radiation domination in the very early universe. Under certain assumptions it might be possible in the future to constrain the dominating equation of state in the early universe via gravitational waves \cite{Boyle_2008}.\\
The scaling behaviour $\varrho_\mrm{SI}\propto{a^{-6}}$ and the proportionality between self-interaction energy density and SIDM particle density $\varrho_\mrm{SI}\propto{n_\mrm{SIDM}^2}$ imply that:
\begin{equation}
 n_\mrm{SIDM}\propto{a^{-3}}
\end{equation}
This condition is naturally fulfilled after DM decoupling and in the case of relativistic particles also before that, as $n_\mrm{WDM}\propto{T^3}$. But it is incompatible with the exponential suppression of non-relativistic particles before they decouple ($n_\mrm{CDM}\propto{\exp{\left(-m_\mrm{CDM}/T\right)}}$). This is why we discuss in the following warm self-interacting dark matter and consider cold dark matter as a second, collisionless component.

We describe the DM as a free Boltzmann gas and additionally take into account a non-zero DM chemical potential. Fermionic DM with a non-vanishing chemical potential according to Fermi-Dirac statistics is analyzed in detail in Ref.\ \cite{Boeckel_2007}. Altogether our model contains six parameters. Besides the particle mass $m_\mrm{WDM}$, the degeneracy factor $g_\mrm{WDM}$ and the relativistic chemical potential $\mu_\mrm{WDM}/T_\mrm{WDM}$, the number of degrees of freedom of particles in thermal equilibrium at WSIDM decoupling $g_\mrm{th\,eq}^\mrm{Wdec}$ (which fixes today's WDM temperature) characterizes the SIDM particle properties. Today's relative amount of WSIDM is given by $F_\mrm{WDM}^0\equiv\Omega_\mrm{WDM}^0/\Omega_\mrm{DM}^0$, and finally $m_\mrm{SI}/\!\sqrt{\alpha_\mrm{SI}}$ determines the WDM self-interaction strength.

\begin{figure}[h]
 \centering
 \includegraphics[width=0.47\textwidth]{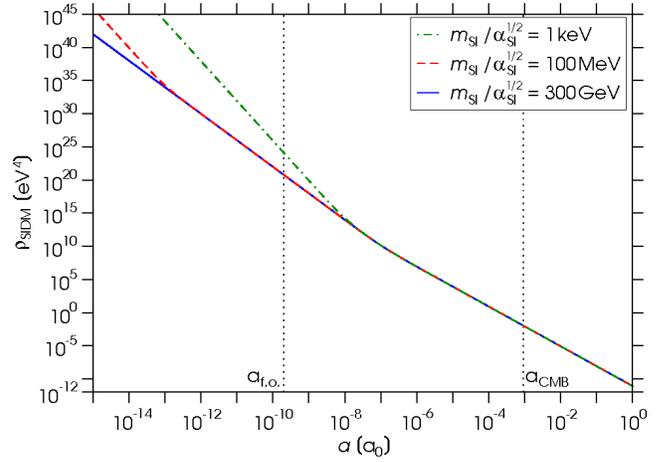}
 \caption[]{Evolution of the SIDM energy density $\varrho_\mrm{SIDM}=\varrho_\mrm{SI}+\varrho_\mrm{WDM}$ with the expansion of the universe for different self-interaction strengths. WDM particle parameters are: $m_\mrm{WDM}=1\,\mrm{keV}$, $g_\mrm{WDM}=2$, $\mu_\mrm{WDM}/T_\mrm{WDM}=0$, $g_\mrm{th\,eq}^\mrm{Wdec}=1107$, $F_\mrm{WDM}^0=1$. $a_\mrm{f.o.}$ is the scale factor at the freeze-out of the neutron to proton number ratio and $a_\mrm{CMB}$ at photon decoupling.}
 \label{fig:rhoSIDM}
\end{figure}

The evolution of the total SIDM energy density $\varrho_\mrm{SIDM}=\varrho_\mrm{SI}+\varrho_\mrm{WDM}$ is shown in Fig.\ \ref{fig:rhoSIDM}. Even very strong DM self-interactions can only have a direct impact on the very early universe. The cosmic microwave background radiation (CMB) is unaffected by the additional DM self-interaction energy density contribution considered. Primordial nucleosynthesis is the major cornerstone to constrain the self-interaction strength. Besides the scaling $\varrho_\mrm{SIDM}\propto{a^{-6}}$ while the self-interaction contribution dominates over the particle contribution, one sees that $\varrho_\mrm{SIDM}\propto{a^{-4}}$ once the relativistic particle contribution dominates and $\varrho_\mrm{SIDM}\propto{a^{-3}}$ when the WDM particles have become non-relativistic.

\section{\label{sec:constraints}constraints}

\subsection{\label{ssec:Omega_DM0}Today's Dark Matter density $\Omega_\mrm{DM}^0$}
The parameters describing the DM particle properties are not all independent of each other, and today's relative DM energy density $\Omega_\mrm{DM}^0$ can be used to extract allowed combinations. At the present-day temperature the WDM particles are non-relativistic, so that their energy density is given by $\varrho_\mrm{WDM}^0=m_\mrm{WDM}\,n_\mrm{WDM}^0$. According to particle number conservation after decoupling and entropy conservation of the particles in thermal equilibrium, today's WDM particle density depends on the one at decoupling as:
\begin{equation}
 \label{eq:nWDM_0dec}
 n_\mrm{WDM}^0=\frac{3\frac{10}{11}}{g_\mrm{th\,eq}^\mrm{Wdec}}\frac{T_0^3}{T_\mrm{Wdec}^3}n_\mrm{WDM}^\mrm{Wdec}
\end{equation}
Since we describe the WDM as a free Boltzmann gas, its number density until decoupling is given by:
\begin{equation}
 \label{eq:n_WDM}
 n_\mrm{WDM}\left(T\right)=\frac{g_\mrm{WDM}}{\pi^2}\,T^3\,\exp\!{\left(\frac{\mu_\mrm{WDM}}{T}\right)}
\end{equation}
Thus Eq.\ (\ref{eq:nWDM_0dec}) leads to the following constraint on the WSIDM particle parameters
\begin{eqnarray}
 \frac{g_\mrm{WDM}\,m_\mrm{WDM}}{g_\mrm{th\,eq}^\mrm{Wdec}\,F_\mrm{WDM}^0}\,\exp\!\left(\frac{\mu_\mrm{WDM}}{T_\mrm{WDM}}\right)&=&\frac{3\pi}{8\times3\frac{10}{11}}\,m_\mrm{Pl}^2\,\frac{H_0^2\,\Omega_\mrm{DM}^0}{T_0^3}\quad\nonumber\\
 &\approx&1.80\,\mrm{eV}\times\frac{\Omega_\mrm{DM}^0h_\mrm{0}^2}{0.1143} \label{eq:Omega0limit}\;,
\end{eqnarray}
with $T_0=({2.725}\pm{0.002})\,\mrm{K}$ \cite{Mather_1999} and $\Omega_\mrm{DM}^0h_0^2=0.1143\pm0.0034$ \cite{Hinshaw_2009}.
\begin{figure}[h]
 \centering
 \includegraphics[width=0.47\textwidth]{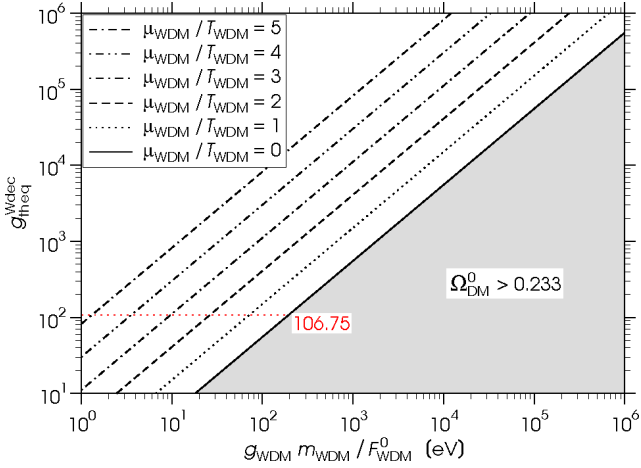}
 \caption[]{Allowed combinations of the parameters describing the WSIDM particle properties according to Eq.\ (\ref{eq:Omega0limit}), so resulting in the correct today's DM energy density $\Omega_\mrm{DM}^0=0.233$.}
 \label{fig:Omega0limits}
\end{figure}
Allowed ranges of the parameters describing the WSIDM particle properties are shown in Fig.\ \ref{fig:Omega0limits} where the degrees of freedom in thermal equilibrium at WSIDM decoupling is plotted against the WDM particle mass over the relative amount of WDM . For the largest possible number of degrees of freedom in thermal equilibrium at WSIDM decoupling in the particle physics standard model, namely $g_\mrm{th\,eq}^\mrm{Wdec}=106.75$, the WDM particle mass is restricted to $g_\mrm{WDM}\,m_\mrm{WDM}\le193\,\mrm{eV}$. To achieve WDM masses of $1-10\,\mrm{keV}$, additional degrees of freedom at WDM decoupling are required. All in all this mass range demands at least $10^3-10^4$ degrees of freedom in thermal equilibrium at WSIDM decoupling. This is common to all WDM models and not specific to the one discussed here. Additional degrees of freedom appear e.g. in supersymmetric theories or in theories with extra dimensions or in string gas cosmology models (see e.g.\ \cite{Brandenberger_2007_1, Brandenberger_2007_2}).

\subsection{\label{ssec:BBN}Primordial nucleosynthesis}
\begin{figure}[h]
 \centering
 \includegraphics[width=0.47\textwidth]{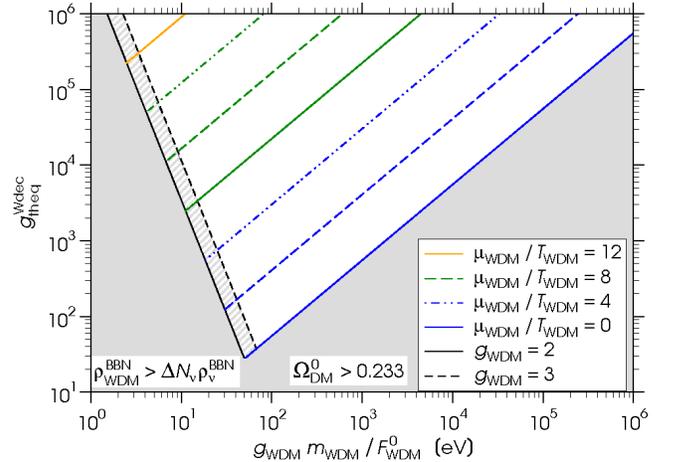}
 \caption[]{Allowed combinations of the WSIDM particle parameters according to Eqs. (\ref{eq:Omega0limit}) and (\ref{eq:DeltaNnu_limit}).}
 \label{fig:mass_window}
\end{figure}

An additional energy density contribution at big bang nucleosynthesis (BBN) is parametrized as additional neutrino families ${\Delta}N_\nu$
\begin{equation}
 \varrho_\mrm{rad}^\mrm{BBN}=\varrho_\gamma^\mrm{BBN}+\left(3+{\Delta}N_\nu\right)\varrho_\nu^\mrm{BBN}+\varrho_\mrm{e}^\mrm{BBN}\;,
\end{equation}
assuming that it scales like radiation ($\varrho_\mrm{rad}\propto{a^{-4}}$). In our model this is true for the energy density contribution of the WDM particles $\varrho_\mrm{WDM}$ but the self-interaction energy density drops faster ($\varrho_\mrm{SI}\propto{a^{-6}}$). Hence, we use limits on ${\Delta}N_\nu$ to constrain $\varrho_\mrm{WDM}$ only ($\varrho_\mrm{WDM}^\mrm{BBN}\le{\Delta}N_\nu\,\varrho_\nu^\mrm{BBN}$), resulting in the following constraint on the WDM particle parameters:
\begin{equation}
 \frac{g_\mrm{WDM}\,\exp\left(\frac{\mu_\mrm{WDM}}{T_\mrm{WDM}}\right)}{{g_\mrm{th\,eq}^\mrm{Wdec}}^{4/3}}\le\frac{7\pi^4}{360\times10.75^{4/3}}\,{\Delta}N_\nu
\end{equation}
Together with Eq.\ (\ref{eq:Omega0limit}) this can be transformed into a lower bound on the WDM particle mass:
\begin{eqnarray}
 \frac{m_\mrm{WDM}}{F_\mrm{WDM}^0}\ge\frac{135}{7\pi^3}\,\frac{10.75^{4/3}}{3\frac{10}{11}}\,m_\mrm{Pl}^2\,\frac{H_0^2\,\Omega_\mrm{DM}^0}{T_0^3}\,\frac{{g_\mrm{th\,eq}^\mrm{Wdec}}^{-1/3}}{{\Delta}N_\nu} \\
 \approx22.6\,\mrm{eV}\times\frac{\Omega_\mrm{DM}^0h_\mrm{0}^2}{0.1143}\,\frac{{g_\mrm{th\,eq}^\mrm{Wdec}}^{-1/3}}{{\Delta}N_\nu} \label{eq:DeltaNnu_limit}
\end{eqnarray}
Ref.\ \cite{Simha_2008} obtains as limit for the additional energy density during BBN (2$\sigma$ uncertainty):
\begin{equation}
 {\Delta}N_\nu\le0.3
\end{equation}
The resulting constraints on the WSIDM particle parameters are shown together with those from today's DM energy density in Fig.\ \ref{fig:mass_window}. While the lower right corner is excluded by today's DM energy density, the lower left one is ruled out by the allowed energy density (which scales like radiation) during BBN. Smaller WDM particle masses require a higher temperature and thus a lower number of degrees of freedom in thermal equilibrium at decoupling to have the correct energy density today. In the grey shaded region at the left the increased WDM temperature would result in a too large energy density during BBN.

\begin{figure}[h]
 \centering
 \includegraphics[width=0.47\textwidth]{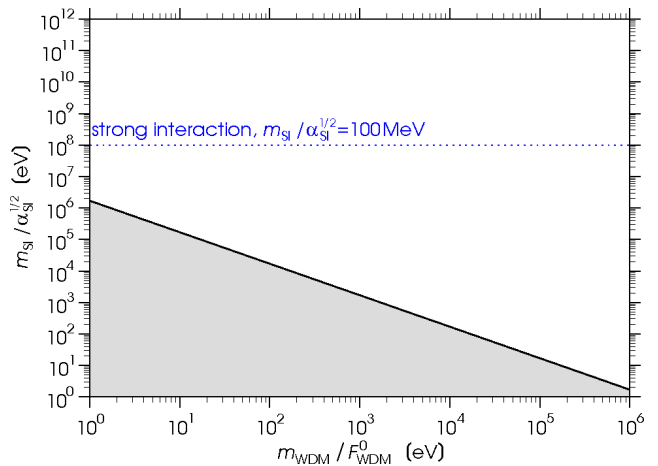}
 \caption[]{Constraint on the DM self-interaction strength according to Eqs. (\ref{eq:maSI_limit}) and (\ref{eq:xSI_limit}) by the permitted energy density at freeze-out of the neutron to proton number ratio as kick-off of the primordial nucleosynthesis.}
 \label{fig:maSI_mDM}
\end{figure}

\begin{figure}[h]
 \centering
 \includegraphics[width=0.47\textwidth]{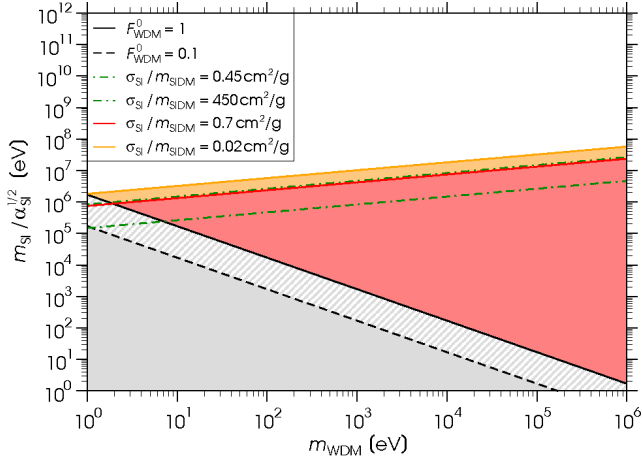}
 \caption[]{Constraint on the DM self-interaction strength depending on the SIDM particle mass for two different relative amounts of SIWDM according to Eqs.~(\ref{eq:maSI_limit}) and (\ref{eq:xSI_limit}), together with the proposed ranges \cite{Spergel_2000, Randall_2008, Miralda_2002} of the DM collisional cross-section, for the case that all DM is self-interacting ($F_\mrm{WDM}^0=1$) according to Eq.~(\ref{eq:sigmaSI})}.
 \label{fig:maSI_comparison}
\end{figure}

To constrain the DM self-interaction strength via BBN we have to focus on the earliest stage of BBN, which is the freeze-out of the neutron to proton number ratio. This occurs at temperatures of $T_\mrm{f.o.}\sim800\,\mrm{keV}$.\\
A larger energy density results in a higher expansion rate of the universe ($H\propto\varrho^{1/2}$). This implies that the equality between expansion rate and reaction rate of reactions that keep the neutrons and protons in equilibrium is achieved earlier, i.e.\ at a higher temperature, and hence the neutron concentration at freeze-out is enhanced. The time when the deuterium bottleneck opens is not affected by an additional energy density of DM self-interactions (as $\varrho_\mrm{SI}\propto{a^{-6}}$). Hence, the period between freeze-out of the relative neutron concentration ($X_\mrm{n}=n_\mrm{n}/\left(n_\mrm{n}+n_\mrm{p}\right)$) and the beginning of nucleosynthesis reactions is slightly extended (${\Delta}t=t_\mrm{bBBN}-t_\mrm{f.o.}$). During this time span free neutrons decay. Nevertheless, also the neutron concentration when the deuterium bottleneck opens is increased ($X_\mrm{n}^\mrm{bBBN}=X_\mrm{n}^\mrm{f.o.}\exp\left(-{\Delta}t/\tau_\mrm{n}\right)$). Nearly all neutrons available for the nucleosynthesis processes are incorporated into $^4\mrm{He}$. For this reason, the primordial $^4\mrm{He}$ mass fraction ($Y_\mrm{P}\simeq2X_\mrm{n}^\mrm{bBBN}$) is the ideal probe to constrain the self-interaction energy density contribution and thus the self-interaction strength. The detailed calculation is given in Appendix \ref{asec:BBN}.\\
We assume for simplicity that the freeze-out of the neutron to proton number ratio occurs in a radiation dominated universe, so that the DM self-interaction energy density contribution does not exceed the radiation contribution:
\begin{equation}
 \varrho_\mrm{tot}^\mrm{f.o.}=\varrho_\mrm{SI}^\mrm{f.o.}+\varrho_\mrm{rad}^\mrm{f.o.}=\left(1+x_\mrm{SI}^\mrm{f.o.}\right)\varrho_\mrm{rad}^\mrm{f.o},\quad0\le{x_\mrm{SI}^\mrm{f.o.}}<1
 \label{eq:rho_tot_fo}\:,
\end{equation}
where $x_\mrm{SI}^\mrm{f.o.}\equiv\varrho_\mrm{SI}^\mrm{f.o.}/\varrho_\mrm{rad}^\mrm{f.o.}$. According to the definition of the DM self-interaction energy density, Eq.\ (\ref{eq:rho_tot_fo}) translates into the following constraint on the DM self-interaction strength:
\begin{eqnarray}
 \frac{m_\mrm{SI}}{\sqrt{\alpha_\mrm{SI}}}=\frac{\sqrt{30}\times10.75}{\sqrt{x_\mrm{SI}^\mrm{f.o}}\,\pi^3}\,\frac{T_\mrm{f.o}}{{g_\mrm{eff}^\mrm{f.o}}^{1/2}}\,\frac{g_\mrm{WDM}}{g_\mrm{th\,eq}^\mrm{Wdec}}\,\exp{\left(\frac{\mu_\mrm{WDM}}{T_\mrm{WDM}}\right)} \nonumber\\
 \label{eq:maSI_limit}
 ={x_\mrm{SI}^\mrm{f.o}}^{-1/2}\,\frac{3\sqrt{30}\times10.75}{8\pi^2\times3\frac{10}{11}}m_\mrm{Pl}^2\frac{T_\mrm{f.o}}{{g_\mrm{eff}^\mrm{f.o}}^{1/2}}\frac{H_0^2\Omega_\mrm{DM}^0}{T_0^3}\frac{F_\mrm{WDM}^0}{m_\mrm{WDM}}\\
 \approx1.70\times10^6\,\mrm{eV^2}\times\frac{\Omega_\mrm{DM}^0h_\mrm{0}^2}{0.1143}\,\frac{T_\mrm{f.o}/879\,\mrm{keV}}{\left(x_\mrm{SI}/0.279\right)^{1/2}}\,\frac{F_\mrm{WDM}^0}{m_\mrm{WDM}} \nonumber
\end{eqnarray}
For the second equality we have used the relation between the parameters describing the WSIDM particle properties according to Eq.\ (\ref{eq:Omega0limit}).\\
A robust upper limit on the primordial $^4\mrm{He}$ abundance inferred from observations is (2$\sigma$ uncertainty, from Ref.~\cite{Steigman_2007}):
\begin{equation}
 Y_\mrm{P}<0.255
\end{equation}
This implies a constraint on the DM self-interaction energy density contribution at the freeze-out of the neutron to proton number ratio (see Appendix \ref{asec:BBN}) as:
\begin{equation}
 \label{eq:xSI_limit}
 x_\mrm{SI}^\mrm{f.o.}<0.279
\end{equation}
The resulting constraint on the DM self-interaction strength according to Eq.\ (\ref{eq:maSI_limit}) is shown in Fig.\ \ref{fig:maSI_mDM}. The grey shaded region is ruled out by the allowed additional energy density contribution of DM self-interactions at the freeze-out of the neutron to proton number ratio. The limit on the self-interaction strength scales inverse proportional to the SIDM particle mass. Even an additional energy density contribution of DM self-interactions of the strength of the strong interaction ($m_\mrm{strong}/\!\sqrt{\alpha_\mrm{strong}}\sim100\,\mrm{MeV}$) is consistent with the primordial element abundances.\\
If we associate with the vector meson exchange interaction a cross-section given by
\begin{equation}
 \label{eq:sigmaSI}
 \sigma_\mrm{SI}\approx{s\,\frac{\alpha_\mrm{SI}^2}{m_\mrm{SI}^4}}\;,
\end{equation}
where $s=4E_\mrm{SIDM}^2$ in the center of momentum frame, with $E_\mrm{SIDM}\sim{T_\mrm{SIDM}}\ \left(\sim{m_\mrm{SIDM}}\right)$ as the relativistic (non-relativistic) single-particle energy, Fig.\ \ref{fig:maSI_comparison} shows the {\textquoteleft}late universe' constraints on the dark matter collisional cross-section as discussed in Sec.\ \ref{sec:introduction}, comparable with the constraint on the DM self-interaction strength from halo structure of Ref.\ \cite{Hogan_2000}.
Note that the constraints on $\sigma_\mrm{SI}/m_\mrm{SIDM}$ are valid only if all DM is self-interacting ($F_\mrm{WDM}^0=1$), whereas our constraint on the DM self-interaction strength via primordial nucleosynthesis has a trivial dependence on the relative amount of SIWDM ($m_\mrm{SI}/\!\sqrt{\alpha_\mrm{SI}}\propto{F_\mrm{WDM}^0}$).

\section{\label{sec:decoupling}Dark Matter decoupling}
Chemical decoupling occurs when the expansion rate of the universe exceeds the dark matter annihilation rate
\begin{equation}
 \Gamma_\mrm{A}=n_\mrm{DM}\,\langle{\sigma_\mrm{A}}v\rangle\;,
\end{equation}
where $\langle{\sigma_\mrm{A}}v\rangle$ is the thermally averaged product of the total DM annihilation cross-section and the relative velocity of the annihilating DM particles. The expansion rate of the universe is determined by the dominant energy density contribution ($H\propto\varrho^{1/2}$). In the standard model DM decoupling takes place in a radiation dominated universe. In an epoch prior to radiation domination when the WDM self-interaction energy density contribution dominates (Fig.\ \ref{fig:rhos_scaling}), DM decoupling can occur during this self-interaction dominated era.

\subsection{\label{ssec:decSIWDM}Self-interacting Warm Dark Matter}
\begin{figure}[h]
 \centering
 \includegraphics[width=0.47\textwidth]{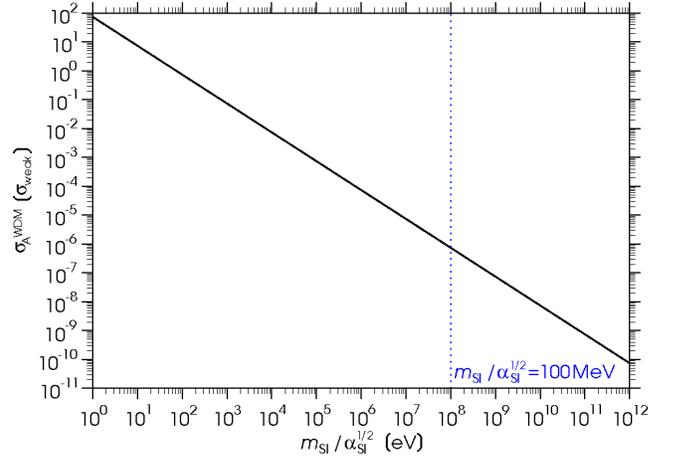}
 \caption[]{Constraint on the WSIDM annihilation cross-section depending on the elastic self-interaction strength, in the case of WSIDM decoupling in a self-interaction dominated universe, according to Eq.\ (\ref{eq:sigmaA_SIWDM}).}
 \label{fig:sigmaA_SIWDM}
\end{figure}

In the case of thermal WDM relics the DM particles are relativistic at decoupling and their annihilation rate is therefore $\Gamma_\mrm{A}^\mrm{WDM}=n_\mrm{WDM}\sigma_\mrm{A}^\mrm{WDM}$. In a universe that is dominated by the WDM self-interaction energy density contribution, also the expansion rate is proportional to the SIWDM particle density ($\varrho_\mrm{SI}\propto{n_\mrm{WDM}^2}$, Eq.\ (\ref{eq:rhoSI})), so that for $\Gamma_\mrm{A}=H$ the WSIDM annihilation cross-section is independent on the particle parameters but determined by the elastic self-interaction strength:
\begin{eqnarray}
 \label{eq:sigmaA_SIWDM}
 \sigma_\mrm{A}^\mrm{WDM}&=&\left(\frac{8\,\pi}{3}\right)^{1/2}m_\mrm{Pl}^{-1}\,\frac{\sqrt{\alpha_\mrm{SI}}}{m_\mrm{SI}}\\
  &\approx&7.45\times10^{-7}\times\frac{100\,\mrm{MeV}}{m_\mrm{SI}/\!\sqrt{\alpha_\mrm{SI}}}\,\sigma_\mrm{weak} \nonumber
\end{eqnarray}
This dependence of the WSIDM annihilation cross-section on the elastic self-interaction strength is shown in Fig.\ \ref{fig:sigmaA_SIWDM}. The annihilation cross-section is given here in units of the cross-section for weak interactions, which is defined as:
\begin{eqnarray}
 \sigma_\mrm{weak}&\equiv&\frac{T_0^3}{m_\mrm{Pl}^3\,H_0^2}\\
 &\approx&3.18\times10^{-12}\,\mrm{GeV^{-2}}\approx1.24\times10^{-39}\,\mrm{cm^2} \nonumber
\end{eqnarray}
One notes that the WSIDM annihilation cross-section is rather small for reasonable (elastic) interaction strengths ($\sigma_\mrm{A}^\mrm{WDM}\ll\sigma_\mrm{weak}$).
Hence, WSIDM decoupling in a self-interaction dominated universe reproduces naturally and consistently the {\textquoteleft}super weak' inelastic coupling between the WSIDM and baryonic matter, required in Sec.\ \ref{ssec:Omega_DM0} to obtain typical WDM particle masses of $1-10\,\mrm{keV}$.\\
Our analysis of SIDM decoupling in a self-interaction dominated universe complies with the qualitative statement of Ref.\ \cite{Hui_2001} ``that the elastic scattering cross section cannot be arbitrarily small given a nonvanishing inelastic cross section''.

\subsection{\label{ssec:decCDM}Collisionless Cold Dark Matter}
\begin{figure}[h]
 \centering
 \includegraphics[width=0.47\textwidth]{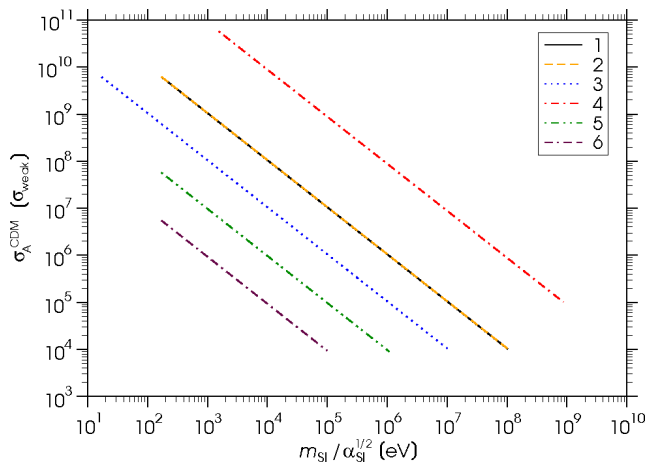}
 \caption[]{Collisionless CDM annihilation cross-section in dependence of the elastic WDM self-interaction strength  according to Eqs.\ (\ref{eq:mT_CDM}) and (\ref{eq:sigmaA_CDM}) for DM particle parameter sets given in Table \ref{tab:DM_parameters}.
 }
 \label{fig:sigmaA_CDM}
\end{figure}

For a CDM species the thermally averaged product of the total DM annihilation cross-section and the relative velocity between the annihilating DM particles is given according to Maxwell-Boltzmann statistics by:
\begin{equation}
 \langle{\sigma_\mrm{A}}v\rangle=\sigma_\mrm{A}^\mrm{CDM}\,\frac{4}{\sqrt{\pi}}\left(\frac{T}{m_\mrm{CDM}}\right)^{1/2}
\end{equation}
Hence, the conditional equation for the decoupling of collisionless CDM in a universe dominated by the self-interaction energy density contribution of SIWDM reads:
\begin{equation}
 \sigma_\mrm{A}^\mrm{CDM}\left(\frac{m_\mrm{CDM}}{T_\mrm{Cdec}}\right)^{-1/2}=\sqrt{\frac{8}{3}}\,\frac{\pi}{4\,m_\mrm{Pl}}\,\frac{n_\mrm{WDM}^\mrm{Cdec}}{n_\mrm{CDM}^\mrm{Cdec}}\,\frac{\sqrt{\alpha_\mrm{SI}}}{m_\mrm{SI}}
\end{equation}
The WDM particle density is that of a decoupled, relativistic free Boltzmann gas. If one inserts for the CDM particle density the particle density of a non-relativistic particle species, one arrives at:
\begin{eqnarray}
 \frac{m_\mrm{CDM}}{T_\mrm{Cdec}}\,\exp{\left(-\frac{m_\mrm{CDM}}{T_\mrm{Cdec}}\right)}&=&\frac{\sqrt{3}\,\pi^{3/2}\,m_\mrm{Pl}}{4\times3\frac{10}{11}}\,\frac{H_0^2\,\Omega_\mrm{DM}^0}{T_0^3}\,\frac{g_\mrm{th\,eq}^\mrm{Cdec}}{\sigma_\mrm{A}^\mrm{CDM}} \nonumber\\
 &&\times\,\frac{F_\mrm{WDM}^0}{g_\mrm{CDM}m_\mrm{WDM}}\,\frac{\sqrt{\alpha_\mrm{SI}}}{m_\mrm{SI}}
 \label{eq:mT_CDM}
\end{eqnarray}
If one takes for the CDM particle density the decoupling particle density according to its today's relic density (Eq.\ (\ref{eq:nWDM_0dec}) for the CDM component), one gets:
\begin{eqnarray}
 \frac{\sigma_\mrm{A}^\mrm{CDM}}{\sigma_\mrm{weak}}&=&8\,\sqrt{\frac{2}{3}}\,m_\mrm{Pl}^2\,\frac{H_0^2\,\Omega_\mrm{DM}^0}{T_0^3}\left(\frac{m_\mrm{CDM}}{T_\mrm{Cdec}}\right)^{1/2} \nonumber\\
 &&\times\,\frac{F_\mrm{WDM}^0}{1-F_\mrm{WDM}^0}\,\frac{m_\mrm{CDM}}{m_\mrm{WDM}}\,\frac{\sqrt{\alpha_\mrm{SI}}}{m_\mrm{SI}}
 \label{eq:sigmaA_CDM}
\end{eqnarray}
While in a radiation dominated universe the CDM annihilation cross-section depends logarithmically on the CDM particle mass, we find that $\sigma_\mrm{A}^\mrm{CDM}$ is proportional to $m_\mrm{CDM}$ when CDM decoupling occurs in a self-interaction dominated universe! Furthermore, a larger elastic self-interaction strength implies a higher expansion rate, so that the annihilation cross-section has to be larger, in order that the CDM fits its particle density today.\\
One can solve Eqs.\ (\ref{eq:mT_CDM}) and (\ref{eq:sigmaA_CDM}) iteratively for a chosen parameter set of DM particle parameters. The correlation between the annihilation cross-section of collisionless CDM and the self-interaction strength of elastic WDM self-interactions is shown in Fig.\ \ref{fig:sigmaA_CDM}.

\begin{table}
 \caption{DM parameter sets shown in Figs.\ \ref{fig:sigmaA_CDM} to \ref{fig:sigmaAv_CDM}. Compared to the reference set 1, we increase the CDM degeneracy factor $g_\mrm{CDM}$ in set 2, the WSIDM particle mass $m_\mrm{WDM}$ in set 3 and the relative amount of WSIDM $F_\mrm{WDM}^0$ in set 4 and decrease the CDM particle mass $m_\mrm{CDM}$ in sets 5 and 6.}
 \begin{ruledtabular}
 \begin{tabular}{lcccc}
  Set & $m_\mrm{CDM}$ & $g_\mrm{CDM}$ & $m_\mrm{WDM}$ & $F_\mrm{WDM}^0$ \\ \hline 
  1 & 10\,TeV & 2 & 1\,keV & 0.1 \\
  2 & 10\,TeV & 3 & 1\,keV & 0.1 \\
  3 & 10\,TeV & 2 & 10\,keV & 0.1 \\
  4 & 10\,TeV & 2 & 1\,keV & 0.9 \\
  5 & 100\,GeV & 2 & 1\,keV & 0.1 \\
  6 & 10\,GeV & 2 & 1\,keV & 0.1 \\
 \end{tabular}
 \end{ruledtabular}
 \label{tab:DM_parameters}
\end{table}

\begin{figure}[h]
 \centering
 \includegraphics[width=0.47\textwidth]{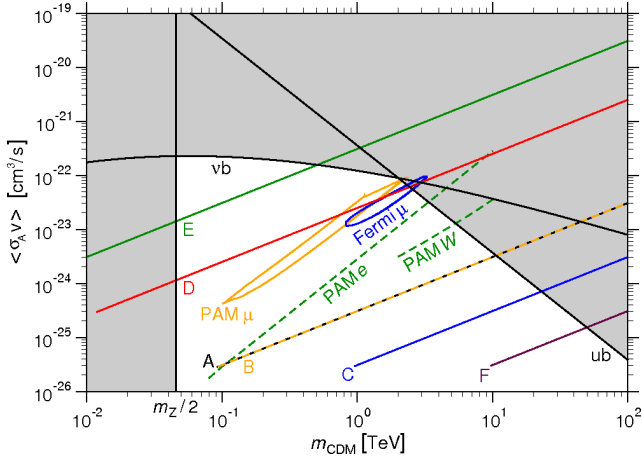}
 \caption[]{Thermally averaged product of the total collisionless CDM annihilation cross-section and the relative velocity between the annihilating CDM particles in dependence of the CDM particle mass for the DM particle parameter sets given in Table \ref{tab:DM_parameters2}, according to Eq.\ (\ref{eq:nat_scale}). Also shown are as upper limits the \emph{Halo Average} neutrino bound of Ref.\ \cite{Yueksel_2007} (${\nu}$b) and the unitarity bound  according to Eq.\ (\ref{eq:ub}) (ub), as well as the $2\sigma$ contours for fits to Fermi (Fermi\,$\mu$) and PAMELA (PAM\,$\mu$) data assuming annihilation only to $\mu^+\mu^-$ of Ref.\ \cite{Bergstroem_2009_B} and the best-fit lines to the PAMELA data for annihilations to $e^+e^-$ (PAM\,$e$) and $W^+W^-$ (PAM\,$W$) of Ref.\ \cite{Catena_2009}.}
 \label{fig:sigmaAv_mCDM}
\end{figure}

The value of $m_\mrm{SI}/\!\sqrt{\alpha_\mrm{SI}}$ is bounded below by constraints from primordial nucleosynthesis according to Eqs.\ (\ref{eq:maSI_limit}) and (\ref{eq:xSI_limit}). The upper limit of $m_\mrm{SI}/\!\sqrt{\alpha_\mrm{SI}}$ comes from requiring a self-interaction dominated universe at CDM decoupling ($\varrho_\mrm{SI}^\mrm{Cdec}>\varrho_\mrm{rad}^\mrm{Cdec}$). This condition implies rather strong elastic WDM self-interactions, which result in large collisionless CDM annihilation cross-sections, exceeding considerably the cross-section of the weak scale. This is in contrast to what is called \textquoteleft{WIMP miracle}' in the standard model, the fact that a CDM particle with a mass around $100\,\mrm{GeV}$ fits to today's DM relic density with an inelastic cross-section of the weak scale. The nonobserved decay of a Z boson into two DM particles rules out CDM masses of $m_\mrm{CDM}\lesssim{m_\mrm{Z}/2}\approx45.6\,\mrm{GeV}$. The linear dependence of the CDM annihilation cross-section on the CDM particle mass can also be recognized in Fig.\ \ref{fig:sigmaA_CDM} (see also Fig.\ \ref{fig:sigmaAv_mCDM}). The ratio of CDM particle mass and temperature at CDM decoupling becomes slightly larger in a self-interaction dominated universe compared to CDM decoupling in a radiation dominated universe, depending on the CDM particle mass. We find values of $m_\mrm{CDM}/T_\mrm{Cdec}$ between $25-35$ for the parameter sets given in Table \ref{tab:DM_parameters}.

\begin{figure}[h]
 \centering
 \includegraphics[width=0.47\textwidth]{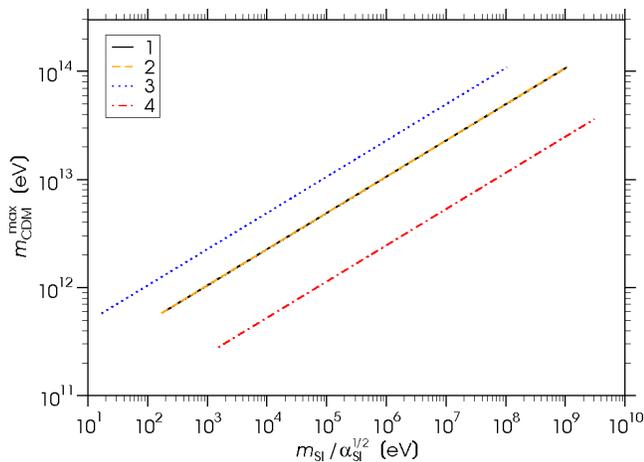}
 \caption[]{Upper limit on the thermal relic CDM particle mass from the unitarity bound depending on the elastic SIWDM self-interaction strength for collisionless CDM decoupling in a self-interaction dominated universe for DM parameter sets 
given in Table \ref{tab:DM_parameters}, according to Eq.\ (\ref{eq:mCDM_max}). }
 \label{fig:mCDM_up}
\end{figure}

\begin{figure}[h]
 \centering
 \includegraphics[width=0.47\textwidth]{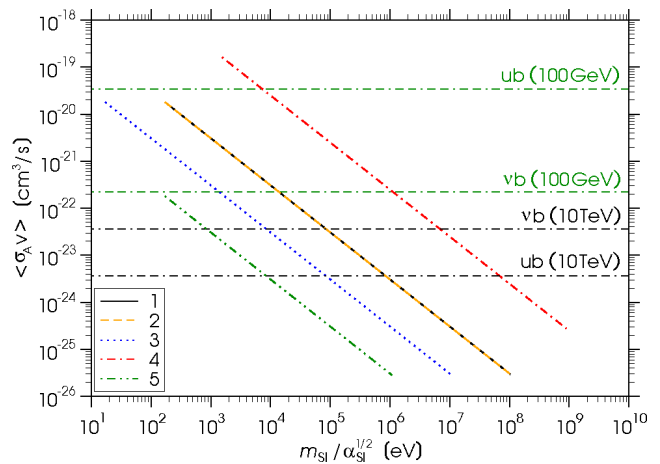}
 \caption[]{Thermally averaged product of the total collisionless CDM annihilation cross-section and the relative velocity between the annihilating CDM particles in dependence of the elastic WDM self-interaction strength for the DM particle parameter sets given in Table \ref{tab:DM_parameters}, according to Eq.\ (\ref{eq:nat_scale}). Also shown are as upper limits the unitarity bound (ub) and neutrino bound (${\nu}$b) for the chosen CDM particle masses.}
 \label{fig:sigmaAv_CDM}
\end{figure}

If CDM decoupling occurs in a radiation dominated universe, the natural scale of the velocity weighted mean annihilation cross-section is $\langle\sigma_\mrm{A}v\rangle\sim3\times10^{-26}\,\mrm{cm^3\,s^{-1}}/\left(1-F_\mrm{WDM}^0\right)$ \cite{Jungman_1996, Beacom_2007}.
For the decoupling of collisionless CDM in a universe dominated by the DM self-interaction energy density contribution this becomes:
\begin{eqnarray}
 \label{eq:nat_scale}
 \langle\sigma_\mrm{A}v\rangle&=&\left(\frac{8\pi}{3}\right)^{1/2}m_\mrm{Pl}^{-1}\,\frac{F_\mrm{WDM}^0}{1-F_\mrm{WDM}^0}\,\frac{m_\mrm{CDM}}{m_\mrm{WDM}}\,\frac{\sqrt{\alpha_\mrm{SI}}}{m_\mrm{SI}}\ \ \ \\
 &\approx&2.77\times10^{-23}\,\mrm{cm^3\,s^{-1}} \nonumber\\
 &&\times\,\frac{m_\mrm{CDM}/10\,\mrm{TeV}}{m_\mrm{WDM}/1\,\mrm{keV}}\,\frac{1\,\mrm{MeV}}{m_\mrm{SI}/\!\sqrt{\alpha_\mrm{SI}}}\,\frac{F_\mrm{WDM}^0}{1-F_\mrm{WDM}^0} \nonumber
\end{eqnarray}
Hence, the degeneracy in the CDM particle mass is removed and the natural scale of the CDM annihilation cross-section depends also on the CDM particle mass. All in all the natural scale of CDM decoupling can be increased by some orders of magnitude when CDM decoupling occurs in a self-interaction dominated universe, depending on the elastic WDM self-interaction strength (see Figs.\ \ref{fig:sigmaAv_CDM} and \ref{fig:sigmaAv_mCDM}). Interestingly enough, such boosted CDM annihilation cross-sections are able to explain the high energy cosmic-ray electron-plus-positron spectrum measured by Fermi-LAT and the excess in the PAMELA data on the positron fraction (e.g.\ \cite{Grasso_2009, Bergstroem_2009_B}, see Fig.\ \ref{fig:sigmaAv_mCDM}).\\
A general upper limit on the DM annihilation cross-section is set by the unitarity bound. For s-wave dominated annihilation this is \cite{Griest_1990, Hui_2001}:
\begin{equation}
 \label{eq:ub}
 \langle\sigma_\mrm{A}v\rangle\le\frac{4\pi}{m_\mrm{CDM}^2\,v}
\end{equation}
Together with the natural scale of the CDM annihilation cross-section, the unitary bound sets an upper limit on the CDM particle mass of thermal relics. For CDM decoupling in a radiation dominated universe this is $m_\mrm{CDM}\lesssim100\,\mrm{TeV}\left(1-F_\mrm{WDM}^0\right)^{1/2}$. If collisionless CDM decoupling occurs in a self-interacting dominated universe, the unitarity bound leads with the corresponding natural scale (Eq.\ (\ref{eq:nat_scale})) to the following limit of the thermal relic CDM particle mass:
\begin{eqnarray}
 m_\mrm{CDM}&\le&\left(\frac{3\pi^2}{8}\right)^{1/6}m_\mrm{Pl}^{1/3}\,\left(\frac{m_\mrm{CDM}}{T_\mrm{Cdec}}\right)^{1/6}\nonumber\\
 &&\times\left(\frac{1-F_\mrm{WDM}^0}{F_\mrm{WDM}^0}\right)^{1/3}m_\mrm{WDM}^{1/3}\left(\frac{m_\mrm{SI}}{\sqrt{\alpha_\mrm{SI}}}\right)^{1/3}
 \label{eq:mCDM_max}\\
 &\approx&2.86\times10^{12}\,\mrm{eV}\times\left(\frac{m_\mrm{WDM}}{1\,\mrm{keV}}\,\frac{m_\mrm{SI}/\!\sqrt{\alpha_\mrm{SI}}}{1\,\mrm{MeV}}\right)^{1/3} \nonumber\\
 &&\times\,\left(\frac{m_\mrm{CDM}}{T_\mrm{Cdec}}\right)^{1/6}\left(\frac{1-F_\mrm{WDM}^0}{F_\mrm{WDM}^0}\right)^{1/3}
\end{eqnarray}
The resulting upper limit on the thermal relic CDM particle mass for CDM decoupling in a self-interaction dominated universe is shown in Fig.\ \ref{fig:mCDM_up}.
The maximum thermal relic CDM particle mass for CDM decoupling in a self-interaction dominated universe depends on the elastic WDM self-interaction strength and on the DM particle parameters. The limit on $m_\mrm{CDM}$ is well in the TeV range.\\
Another way to constrain the annihilation of DM is via the appearance of thereby produced particles. Neutrinos proved to be the most useful final state, since they provide a stringent but conservative upper limit on the DM annihilation cross-section independent of branching ratios \cite{Beacom_2007, Yueksel_2007}. Assuming s-wave dominated CDM annihilation processes we can directly transfer the DM annihilation cross-section limits from today's neutrino signal to the early universe CDM decoupling. Since our model contains three ingredients, namely, warm dark matter, non-vanishing elastic dark matter self-interactions, and larger cold dark matter annihilation cross-sections, which tend to lead to less cuspy halo profiles, we use the Milky Way \emph{Halo Average} neutrino constraint of Ref.\ \cite{Yueksel_2007} to compare with the predicted CDM annihilation cross-sections when collisionless CDM decoupling occurs in a self-interaction dominated universe. Fig.\ \ref{fig:sigmaAv_CDM} shows the velocity weighted mean CDM annihilation cross-section for decoupling in a self-interaction dominated universe according to Eq.\ (\ref{eq:nat_scale}) together with the unitarity bound and neutrino bound for given CDM particle mass. One realizes that the combination of superstrong WDM self-interaction strengths together with very heavy CDM particle masses are ruled out. With regard to the neutrino bound of the DM annihilation cross-section one should consider that it assumes that all the DM is collisionless CDM, which is not the case in our model with a much more weakly annihilating WDM component, so that it sets very strong limits on the self-interaction strengths.

Finally Fig.\ \ref{fig:sigmaAv_mCDM} displays the velocity weighted mean CDM annihilation cross-section for decoupling in a self-interaction dominated universe according to Eq.\ (\ref{eq:nat_scale}) in the common presentation as a function of the collisionless CDM particle mass, together with the \emph{Halo Average} neutrino constraint of Ref.\ \cite{Yueksel_2007}, the unitary bound according to Eq.\ (\ref{eq:ub}), the $2\sigma$ contours for fits to Fermi and PAMELA data assuming annihilation only to $\mu^+\mu^-$ of Ref.\ \cite{Bergstroem_2009_B}, and the best-fit lines to the PAMELA data for annihilations to $e^+e^-$ and $W^+W^-$, respectively, of Ref.\ \cite{Catena_2009}.

\begin{table}
 \caption{DM parameter sets shown in Fig.\ \ref{fig:sigmaAv_mCDM}. Compared to the reference set A, we increase the CDM degeneracy factor $g_\mrm{CDM}$ in set B, the WSIDM particle mass $m_\mrm{WDM}$ in set C, the relative amount of WSIDM $F_\mrm{WDM}^0$ in set D and the elastic WDM self-interaction strength in set E and decrease it in set F.}
 \begin{ruledtabular}
 \begin{tabular}{lcccc}
  Set & $m_\mrm{SI}/\!\sqrt{\alpha_\mrm{SI}}$ & $g_\mrm{CDM}$ & $m_\mrm{WDM}$ & $F_\mrm{WDM}^0$ \\ \hline 
  A & 1\,MeV & 2 & 1\,keV & 0.1 \\
  B & 1\,MeV & 3 & 1\,keV & 0.1 \\
  C & 1\,MeV & 2 & 10\,keV & 0.1 \\
  D & 1\,MeV & 2 & 1\,keV & 0.9 \\
  E & 1\,keV & 2 & 1\,keV & 0.1 \\
  F & 100\,MeV & 2 & 1\,keV & 0.1 \\
 \end{tabular}
 \end{ruledtabular}
 \label{tab:DM_parameters2}
\end{table}

\section{\label{sec:struct formation}Structure formation}

Now we want to study the impact of a self-interaction-dominated epoch in the early universe on the evolution of ideal fluid cosmological perturbations. 
We use uniform expansion gauge (UEG) that is free of unphysical gauge modes and has the two gauge invariant variables $\delta$ and $\hat\psi$, which can be identified with the density contrast and a quantity related to the fluid velocity in the subhorizon limit ($k_{\mrm{ph}} \gg H$), respectively \cite{Schmid_1999,Hwang_1993}.

The system of general relativistic evolution equations in UEG reads
\begin{eqnarray} 
\delta_i^\prime&=&\frac{3\left(w_i-c^2_{\mrm{s}i}\right)}{a}\delta_i+\frac{k}{a\mathcal{H}}\hat\psi_i-\frac{3\left(1+w_i\right)}{a}\alpha\label{delta}\\
\hat\psi_i^\prime&=&\frac{3w_i-1}{a}\hat\psi_i-c^2_{\mrm{s}i}\frac{k}{a\mathcal{H}}\delta_i-\frac{\left(1+w_i\right)k}{a\mathcal{H}}\alpha\label{psi}\\
\label{alpha}\alpha&=&-\frac{\frac{3}{2}\left(1+3c^2_\mrm{s}\right)}{\left(\frac{k}{\mathcal{H}}\right)^2+\frac{9}{2}\left(1+w\right)}\delta\;,
\end{eqnarray}
where $\delta$ is the density contrast, $w = p/\varrho$ is the equation of state, $c_\mrm{s}$ is the isentropic speed of sound, $\mathcal{H}$ is the conformal Hubble parameter, $k$ is the comoving wavenumber of the mode ($k=k_\mrm{ph}\,a$), and $\alpha$ is the perturbation of the lapse. Primes denote derivatives with respect to the scale parameter $a$.
Eqs.\ (\ref{delta}) and (\ref{psi}) apply to each decoupled fluid component $i$ individually and the general relativistic analogue of the Poisson equation (\ref{alpha}) connects them. In Eq.\ (\ref{alpha}) the averaged quantities $w = \sum_i p_i / \sum_i \varrho_i$, $\delta = \sum_i (\varrho_i \delta_i) / \sum_i \varrho_i $ and $ c^2_\mrm{s}  = \sum_i (c^2_{\mrm{s}i} \varrho_i \delta_i) / \sum_i (\varrho_i \delta_i) $ enter.\\
For a single fluid with $w = 1$, corresponding according to Eq.\ (\ref{eq:rhoSI}) to self-interaction domination, the analytic solution in the subhorizon limit (${k}/{\mathcal{H}} \gg 1$) is given by
\begin{equation}
\delta_{\mrm{SI}}(a) \propto a \cdot \left(A \cos(a^2 - 3\pi/4) + B \sin(a^2 - 3\pi/4)\right) \label{deltaSI}\;,
\end{equation}
i.e.\ an oscillation with linearly growing amplitude as shown by Ref.\ \cite{Hwang_1993}. This is in contrast to density fluctuations in standard CDM that can only grow logarithmically during radiation domination in the early universe.

For self-interacting warm dark matter there are two relevant damping scales, collisional self-damping (sd) due to particle scattering at early times and free streaming (fs) at late times when the WDM elastic self-interaction rate $\Gamma_{\mrm{SI}}$ has dropped below the Hubble rate. These two can be estimated via the expressions
\begin{eqnarray}
l_{\mrm{sd}}^2&\approx&\int^{t_{\mrm{sdec}}}_0\frac{v_{\mrm{WDM}}^2(t)\,dt}{\Gamma_{\mrm{SI}}(t)\,a^2(t)}\label{damp}\\
l_{\mrm{fs}}&\approx&\int^{t_{\mrm{collapse}}}_{t_{\mrm{sdec}}}\frac{v_{\mrm{WDM}}(t)\,dt}{a(t)}
\end{eqnarray}
as given, for example, in Refs.\ \cite{Boehm_2005, Boehm_2002}. $t_{\mrm{collapse}}$ denotes the time of gravitational collapse, and the rms velocity of the WDM particles is $v_\mrm{WDM}=c=1$. Eq.\ (\ref{damp}) is only valid as long as $l_{\mrm{sd}} \ll l_{\mrm{fs}}$, i.e.\ as long as the particles can be treated as interacting. Any increase in the density contrast in WSIDM produced at early times will be washed out at later times either due to collisional self-damping or due to an inevitable phase of free streaming after self-decoupling (sdec).

In a mixed model of SIWDM and collisionless CDM the picture can differ because the CDM component allows some increase in density fluctuations to be stored. First of all let us examine the solution to the Eqs.\ (\ref{delta}) to (\ref{alpha}) for a subdominant CDM component in a SIWDM background ($w_\mrm{CDM}=c^2_\mrm{s\,CDM}=0, w\approx{w_\mrm{SI}}=1, c^2_\mrm{s}\approx{c^2_\mrm{s\,SI}}=1$):
\begin{eqnarray} 
\delta_{\mrm{CDM}}^\prime&=&\frac{k}{a\mathcal{H}}\hat\psi_{\mrm{CDM}}-\frac{3}{a}\alpha\label{deltaCDM}\\
\hat\psi_{\mrm{CDM}}^\prime&=&-\frac{1}{a}\hat\psi_{\mrm{CDM}}-\frac{k}{a\mathcal{H}}\alpha\label{psiCDM}\\
\label{alphaCDM} \alpha&=&-\frac{6}{\left(\frac{k}{\mathcal{H}}\right)^2+9}\delta_{\mrm{SI}}
\end{eqnarray}
The transition from superhorizon to subhorizon behavior happens very quickly in a SIDM background since ${k}/{\mathcal{H}}={k_{\mrm{ph}}}/{H}= \left({a}/{a^{\mrm{in}}_{k}}\right)^2$, where $a^{\mrm{in}}_{k}$ is the scale parameter at horizon entry. The terms proportional to $\alpha$ can be dropped in the subhorizon limit (${k}/{\mathcal{H}}\gg1$) and the equations simplify further:
\begin{eqnarray} 
\delta_{\mrm{CDM}}^\prime&=&\frac{a}{{a^\mrm{in}_{k}}^2}\hat\psi_{\mrm{CDM}}\label{deltaCDM1}\\
\hat\psi_{\mrm{CDM}}^\prime&=&-\frac{1}{a}\hat\psi_{\mrm{CDM}}\label{psiCDM1}
\end{eqnarray}
The solution for $\hat\psi_{\mrm{CDM}}$ is easily found to be $\hat\psi_{\mrm{CDM}}= C/a$ with $C$ being a constant. This then automatically yields the solution to Eq.\ (\ref{deltaCDM1})
\begin{equation}
 \delta_{\mrm{CDM}} = a \cdot \left(C/{a^\mrm{in}_{k}}^2\right) + D\;,
\end{equation}
with $D$ being a constant. Interestingly, this means that subhorizon collisionless CDM density fluctuations will also grow linearly during a SIWDM dominated phase in contrast to a radiation dominated phase. Thus there will be a region of enhanced fluctuations at low masses in the matter power spectrum between the comoving wavenumber that is equal to the Hubble scale $\mathcal{H}^\mrm{eq}$ at SIDM-radiation equality  and the wavenumber that corresponds to the collisional self-damping scale $k^{\mrm{eq}}_{\mrm{sd}}$ of SIDM at the same moment. A quick estimate then yields the following CDM transfer function $T(k) = A^{\mrm{eq}}(k)/A^{\mrm{in}}(k)$ i.e.\ the ratio of the amplitude of the CDM density fluctuation $A$ with wavenumber $k$ at SIDM-radiation equality (eq) normalized to the amplitude  at horizon crossing (in):
\begin{equation}
T(k) = \sqrt{\frac{k}{\mathcal{H}^{\mrm{eq}}}}\;,\quad k^{\mrm{eq}}_{\mrm{sd}} > k > \mathcal{H}^\mrm{eq}\label{transfer}
\end{equation}
This results from the fact that in a SIDM background modes become subhorizon more quickly since ${k}/{\mathcal{H}}\propto a^2$ while each subhorizon CDM mode only grows as $\delta_{\mrm{CDM}} \propto a$. So while each subhorizon mode has increased in amplitude by one order of magnitude, two additional orders of magnitude in wavenumber have become subhorizon. Therefore the spectrum is less steep than one might expect.

\begin{figure}[h]
 \centering
 \includegraphics[width=0.47\textwidth]{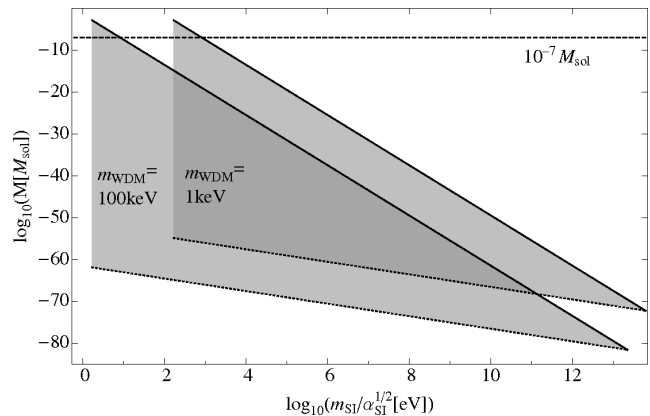}
 \caption[]{Mass range of collisionless CDM density fluctuations affected by a SIWDM dominated epoch for two values of the SIWDM dark matter mass as a function of the self-interaction strength at $\mu_\mrm{WDM}/T_\mrm{WDM}=0$ and $F_\mrm{WDM}^0=0.1$.}
 \label{fig:MaxMass}
\end{figure}

In Refs.\ \cite{Alcock_1998, Afonso_2003} limits on the abundance of planet size dark matter objects in the galactic halo via gravitational lensing from the MACHO and EROS surveys are given. These surveys are sensitive to dark matter objects down to $\sim10^{-7} M_\odot$. In Fig.\ \ref{fig:MaxMass} we show the affected mass range of collisionless CDM density fluctuations as a function of the WDM elastic self-interaction strength $m_{\mrm{SI}}/\!\sqrt{\alpha_\mrm{SI}}$ for two different values of the SIWDM particle mass $m_\mrm{WDM} = 1\,\mrm{keV}, 100\,\mrm{keV}$ at $\mu_\mrm{WDM}/T_\mrm{WDM}=0$ and $F_\mrm{WDM}^0=0.1$. Also shown is the sensitivity limit of the MACHO and EROS surveys. The shaded areas are limited to the left by the requirement $a_{\mrm{eq}}^\mrm{SI\,rad} < a_{\mrm{BBN}}$ (Eqs.\ (\ref{eq:maSI_limit}) and (\ref{eq:xSI_limit})) which also limits the largest structures that can be affected to $\sim1.4 \times 10^{-3} M_\odot$. Using Eq.\ (\ref{transfer}) this means in turn that fluctuations on scales of $10^{-7} M_\odot$ are only enhanced by at most a factor of $\sim\left({10^{-3}}/{10^{-7}}\right)^{1/6}\approx 5$. Therefore it is unlikely that an observable overproduction of planet sized objects could result from the proposed scenario. For most of the parameter space the effect is far from being observable with present small scale dark matter surveys. In the limit of very weak WDM self-interactions the self-damping length becomes very large and WSIDM can become free streaming before SIDM-radiation equality and any temporary small scale increase in the density contrast gets damped away.\\
Note that elastic scattering processes between collisionless CDM and standard model particles can contribute an additional induced collisional damping scale until CDM thermal decoupling, discussed e.g.\ in Refs.\ \cite{Boehm_2005, Schmid_1999, Bringmann_2007}.

\section{\label{sec:conclusions}Conclusions}
In this paper we have analysed constraints on an energy density contribution of elastic dark matter self-interactions $\varrho_\mrm{SI}$, characterized by the mass of the exchanged particle $m_\mrm{SI}$ and the coupling constant $\alpha_\mrm{SI}$.\\
The scaling of energy densities implied that the self-interaction contribution decreases as $\varrho_\mrm{SI}\propto{a^{-6}}$ and thus can only have a direct impact on the very early universe. As the energy density scales with the number density squared due to interactions, self-interacting dark matter has to be warm in the case of thermal relics to give the correct scaling behaviour $n_\mrm{SIDM}\propto{a^{-3}}$. Note that this does not rule out a second collisionless cold dark matter component.\\
We used today's dark matter energy density and the allowed radiation energy density during primordial nucleosynthesis to constrain the parameters characterising the warm self-interacting dark matter particle properties. The dependence of the primordial $^4\mrm{He}$ abundance on the dark matter self-interaction energy density contribution at neutron to proton number ratio freeze-out allowed to constrain the self-interaction strength $m_\mrm{SI}/\!\sqrt{\alpha_\mrm{SI}}$, which depends inversely on the self-interacting dark matter particle mass ($m_\mrm{SI}/\!\sqrt{\alpha_\mrm{SI}}\propto{1/m_\mrm{WDM}}$) but can be at least as strong as the strong interaction scale ($m_\mrm{SI}/\!\sqrt{\alpha_\mrm{SI}}\sim100\,\mrm{MeV}$). Furthermore, our constraint on the dark matter self-interaction strength has a trivial dependence on the relative amount of self-interacting warm dark matter ($m_\mrm{SI}/\!\sqrt{\alpha_\mrm{SI}}\propto{F_\mrm{WDM}^0}$).

We also analyzed dark matter decoupling in a universe dominated by the self-interaction energy density contribution. The annihilation cross-section of warm self-interacting dark matter $\sigma_\mrm{A}^\mrm{WDM}$ is inverse proportional to the elastic self-interaction strength ($\sigma_\mrm{A}^\mrm{WDM}\propto\sqrt{\alpha_\mrm{SI}}/m_\mrm{SI}$) and much smaller than $\sigma_\mrm{weak}$. The natural scale for the annihilation cross-section of a collisionless cold dark matter component $\sigma_\mrm{A}^\mrm{CDM}$ exceeds the weak scale ($\sigma_\mrm{A}^\mrm{CDM}>\sigma_\mrm{weak}$) and depends linearly on the particle mass $m_\mrm{CDM}$ ($\sigma_\mrm{A}^\mrm{CDM}\propto{m_\mrm{CDM}\times\sqrt{\alpha_\mrm{SI}}/m_\mrm{SI}}$). This casts new light on the {\textquoteleft}WIMP miracle' and coincides with the Fermi-LAT and PAMELA data. The unitary bound and neutrino induced constraints on the dark matter annihilation cross-section allowed to disfavour the combination of superstrong elastic warm dark matter self-interactions ($m_\mrm{SI}/\!\sqrt{\alpha_\mrm{SI}}\lesssim1\,\mrm{MeV}$) together with very heavy thermal relic cold dark matter particle masses ($m_\mrm{CDM}\sim10\,\mrm{TeV}$).\\
A relativistic analysis of linear perturbation theory reveals a linear growing solution $\delta\propto{a}$ of self-interaction dominated warm dark matter and also of collisionless cold dark matter in a mixed model during self-interaction domination. However, only non-cosmological scales ($M\lesssim10^{-3}M_\odot$) can be enhanced and a small observable effect could only be present with fine-tuned parameters.

\begin{acknowledgments}
We thank Andrea Macci\`{o} for discussions about warm dark matter particle masses, Hasan Y\"{u}ksel and John Beacom for providing us with their \emph{Halo Average} data of Ref.\ \cite{Yueksel_2007}, Gabrijela Zaharijas for providing us with the $2\sigma$ contours for fits to Fermi and PAMELA data assuming annihilation only to $\mu^+\mu^-$ of Ref.\ \cite{Bergstroem_2009_B}, and Riccardo Catena for providing us with the best-fit lines to the PAMELA data for annihilations to $e^+e^-$ and $W^+W^-$ of Ref.\ \cite{Catena_2009}.\\
This work was supported by the German Research Foundation (DFG) within the framework of the excellence initiative through the Heidelberg Graduate School of Fundamental Physics (HGSFP) and through the Graduate Program for Hadron and Ion Research (GP-HIR) by the Gesellschaft f\"ur Schwerionenforschung (GSI), Darmstadt.
\end{acknowledgments}

\appendix

\section{\label{asec:alphaSI}Self-interaction coupling constant}
\begin{figure}[h]
 \centering
 \includegraphics[width=0.47\textwidth]{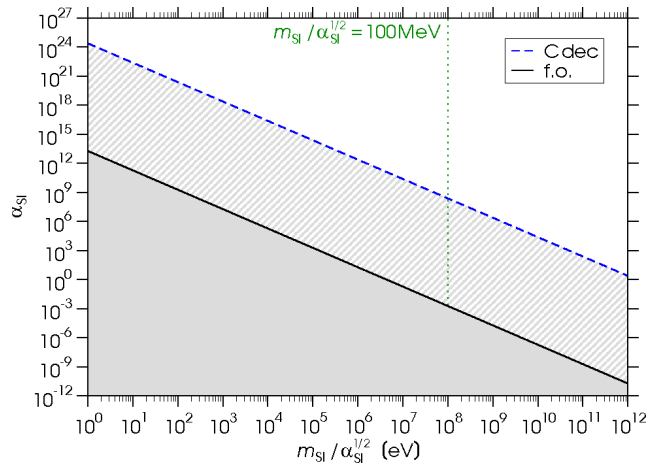}
 \caption[]{Constraints on the coupling constant in dependence of the DM self-interaction strength at primordial nucleosynthesis and collisionless CDM decoupling, according to Eq.\ (\ref{eq:alphaSI}).}
 \label{fig:alphaSI}
\end{figure}

The ansatz for the self-interaction energy density that we have used in this work (Eq.\ \ref{eq:rhoSI}) is valid only when $m_\mrm{SI}>5\,T_\mrm{SIDM}$. Once $m_\mrm{SI}/\!\sqrt{\alpha_\mrm{SI}}$ is known, this can also be used as a boundary condition on the coupling constant for given $T_\mrm{SIDM}$:
\begin{equation}
 \label{eq:alphaSI}
 \alpha_\mrm{SI}>25\,\frac{\alpha_\mrm{SI}}{m_\mrm{SI}^2}\,T_\mrm{SIDM}^2
\end{equation}
This condition has to be fulfilled at primordial nucleosynthesis and at collisionless CDM decoupling for the two component DM scenario. Fig.\ \ref{fig:alphaSI} shows the corresponding constraints on the self-interaction coupling constant for the conservative assumptions $T_\mrm{WDM}^\mrm{f.o.}=T_\mrm{f.o.}$ and $T_\mrm{WDM}^\mrm{Cdec}=m_\mrm{CDM}/\left(m_\mrm{CDM}/T_\mrm{Cdec}\right)$ with $m_\mrm{CDM}=10\,\mrm{TeV}$.
We see that a self-interaction dominated universe at BBN imposes rather loose conditions on the coupling constant (e.g.\ $\alpha_\mrm{strong}>1.93\times10^{-3}$), while collisionless CDM decoupling requires exotic coupling constants (e.g.\ $\alpha_\mrm{strong}>2.47\times10^{8}$).

\section{\label{asec:BBN}Primordial nucleosynthesis}
\begin{figure}[h]
 \centering
 \includegraphics[width=0.47\textwidth]{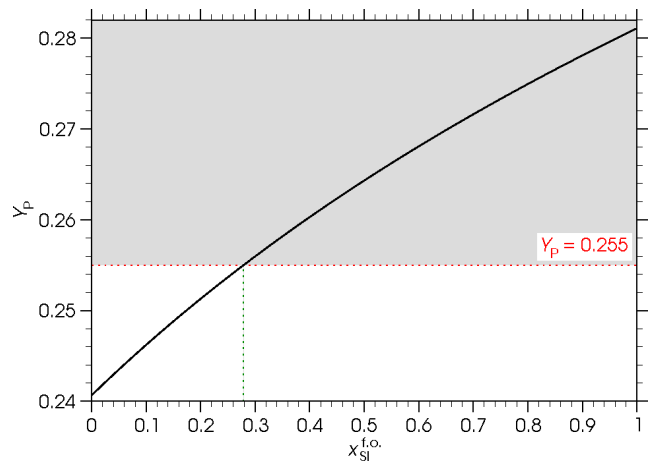}
 \caption[]{Primordial $^4\mrm{He}$ mass abundance $Y_\mrm{p}$ depending on the additional energy density of DM self-interactions at neutron to proton number ratio freeze-out. Additionally the upper limit on $Y_\mrm{p}$ inferred from observations is shown.}
 \label{fig:YP}
\end{figure}

The following analytical calculations are based on Refs.\ \cite{Mukhanov_2004, Mukhanov_2005}.
We assume that the freeze-out of the neutron to proton number ratio occurs in a radiation dominated universe, so that the DM self-interaction energy density contribution does not exceed the radiation contribution:
\[
 \varrho_\mrm{tot}^\mrm{f.o.}=\varrho_\mrm{SI}^\mrm{f.o.}+\varrho_\mrm{rad}^\mrm{f.o.}=\left(1+x_\mrm{SI}^\mrm{f.o.}\right)\varrho_\mrm{rad}^\mrm{f.o},\quad0\le{x_\mrm{SI}^\mrm{f.o.}}<1
\]
The modification of the total energy density implies a change in the temperature-time relation, and thus the conditional equation of the neutron to proton ratio freeze-out temperature becomes:
\begin{eqnarray}
 9.50\!\left(\frac{T_\mrm{f.o.}}{{\Delta}m}\right)^4+4.63\!\left(\frac{T_\mrm{f.o.}}{{\Delta}m}\right)^3+0.677\!\left(\frac{T_\mrm{f.o.}}{{\Delta}m}\right)^2=\nonumber\\
 ={g_{\mrm{eff}}^\mrm{f.o.}}^{1/2}\left(1+x_\mrm{SI}^\mrm{f.o.}\right)^{1/2}\ 
\end{eqnarray}
$g_\mrm{eff}^\mrm{f.o.}$ is the effective number of degrees of freedom contributing to the radiation energy density. According to the discussion in Sec.\ \ref{ssec:BBN} $g_\mrm{eff}^\mrm{f.o.}=11.275$. Adopting the largest possible energy density of radiation during BBN, we calculate the most conservative limits on the self-interaction energy density and hence self-interaction strength. The freeze-out temperature without any self-interaction energy density contribution -- but with the maximum WDM particle contribution $\varrho_\mrm{WDM}^\mrm{f.o.}=0.3\,\varrho_\nu^\mrm{f.o.}$ -- is $T_\mrm{f.o.}(x_\mrm{SI}^\mrm{f.o.}=0)\approx848\,\mrm{keV}$. Doubling the radiation energy density at neutron to proton number ratio freeze-out, the freeze-out temperature would increase to $938\,\mrm{keV}$.
The increase of the freeze-out temperature by a non-vanishing DM self-interaction energy density contribution at freeze-out results in an increase of the relative neutron concentration at freeze-out, which is given by:
\begin{eqnarray}
 X_\mrm{n}^{\mrm{f.o.}}&=&\int\limits_0^{\infty}\exp\Bigg\{-{g_\mrm{eff}}^{-{1/2}}\left(1+x_\mrm{SI}\right)^{-{1/2}}\int\limits_0^y\Big[9.50x^2\nonumber\\
 &&+\,4.63x+0.677\Big]\left[1+\exp\!\left(-\frac{1}{x}\right)\right]\mrm{d}x\Bigg\}\nonumber\\
 &&\times\,\frac{\mrm{d}y}{2y^2\left[1+\cosh\left(\frac{1}{y}\right)\right]}
\end{eqnarray}
For a vanishing energy density contribution of DM self-interactions the relative freeze-out neutron concentration is $X_\mrm{n}^\mrm{f.o.}(x_\mrm{SI}^\mrm{f.o.}=0)\approx0.157$ which would increase to 0.184 if one doubles the radiation energy density. The number of neutrons available for the primordial nucleosynthesis processes depends on the time spent between the freeze-out and the opening of the deuterium bottleneck. The moment of the neutron concentration freeze-out (using the point in time corresponding to the above defined freeze-out temperature) in a radiation dominated universe is:
\begin{equation}
 t_\mrm{f.o.}=\left(\frac{45}{16\,\pi^3}\right)^{1/2}m_\mrm{Pl}\,{g_\mrm{eff}^\mrm{f.o.}}^{-1/2}\left(1+x_\mrm{SI}^\mrm{f.o.}\right)^{-1/2}T_\mrm{f.o.}^{-2}
\end{equation}
The time corresponding to the temperature when the nucleosynthesis processes effectively set in is given by
\begin{equation}
  t_\mrm{bBBN}=\left(\frac{45}{16\,\pi^3}\right)^{1/2}m_\mrm{Pl}\,{g_\mrm{eff}^\mrm{bBBN}}^{-1/2}\,T_\mrm{bBBN}^{-2}
\end{equation}
since the energy density contribution of the self-interaction is vanishing by then (see the discussion in Sec.\ \ref{ssec:BBN}). Because $T_\mrm{bBBN}\simeq73.7\,\mrm{keV}$ the corresponding effective number of relativistic degrees of freedom is $g_\mrm{eff}^\mrm{bBBN}\approx3.50$, where we include the WDM particles as being still relativistic. For relatively heavy WDM particles that decouple at a large number of degrees of freedom in thermal equilibrium this does not necessarily be the case, but it is again the right choice for a conservative constraint of the self-interaction strength.
With this input one can calculate the relative neutron concentration at the effective beginning of the nucleosynthesis ($X_\mrm{n}^\mrm{bBBN}=X_\mrm{n}^\mrm{f.o.}\exp\left(-{\Delta}t/\tau_\mrm{n}\right)$) and finally the expected primordial abundance of $^4\mrm{He}$ ($Y_\mrm{P}\simeq2X_\mrm{n}^\mrm{bBBN}$).
The dependence of the primordial $^4\mrm{He}$ abundance on an additional energy density contribution of DM self-interactions at freeze-out of the number ratio of neutrons and protons is shown in Fig.\ \ref{fig:YP}. Without this contribution it is $Y_\mrm{P}(x_\mrm{SI}^\mrm{f.o.}=0)\approx0.241$ and in the case of twice the radiation energy density this value increases to 0.281.\\
The primordial $^4\mrm{He}$ abundance inferred from observations is subject of systematic uncertainties (for a discussion see Ref.\ \cite{Steigman_2007}) but a robust upper limit on $Y_\mrm{P}$ is $Y_\mrm{P}<0.255$ (2$\sigma$, \cite{Steigman_2007}). This implies a constraint on the DM self-interaction energy density contribution at the freeze-out of the neutron to proton number ratio of $x_\mrm{SI}^\mrm{f.o.}<0.279$.\\
Very recently Refs.\ \cite{Aver_2010, Izotov_2010} determined the primordial $^4\mrm{He}$ abundance $Y_\mrm{P}$ with a central value of $0.256$, which underlines the possibility of new physics beyond standard BBN.

\bibliography{citations}

\begin{thebibliography}{64}
\expandafter\ifx\csname natexlab\endcsname\relax\def\natexlab#1{#1}\fi
\expandafter\ifx\csname bibnamefont\endcsname\relax
  \def\bibnamefont#1{#1}\fi
\expandafter\ifx\csname bibfnamefont\endcsname\relax
  \def\bibfnamefont#1{#1}\fi
\expandafter\ifx\csname citenamefont\endcsname\relax
  \def\citenamefont#1{#1}\fi
\expandafter\ifx\csname url\endcsname\relax
  \def\url#1{\texttt{#1}}\fi
\expandafter\ifx\csname urlprefix\endcsname\relax\def\urlprefix{URL }\fi
\providecommand{\bibinfo}[2]{#2}
\providecommand{\eprint}[2][]{\url{#2}}

\bibitem[{\citenamefont{{Hinshaw} et~al.}(2009)\citenamefont{{Hinshaw},
  {Weiland}, {Hill}, {Odegard}, {Larson}, {Bennett}, {Dunkley}, {Gold},
  {Greason}, {Jarosik} et~al.}}]{Hinshaw_2009}
\bibinfo{author}{\bibfnamefont{G.}~\bibnamefont{{Hinshaw}}},
  \bibinfo{author}{\bibfnamefont{J.~L.} \bibnamefont{{Weiland}}},
  \bibinfo{author}{\bibfnamefont{R.~S.} \bibnamefont{{Hill}}},
  \bibinfo{author}{\bibfnamefont{N.}~\bibnamefont{{Odegard}}},
  \bibinfo{author}{\bibfnamefont{D.}~\bibnamefont{{Larson}}},
  \bibinfo{author}{\bibfnamefont{C.~L.} \bibnamefont{{Bennett}}},
  \bibinfo{author}{\bibfnamefont{J.}~\bibnamefont{{Dunkley}}},
  \bibinfo{author}{\bibfnamefont{B.}~\bibnamefont{{Gold}}},
  \bibinfo{author}{\bibfnamefont{M.~R.} \bibnamefont{{Greason}}},
  \bibinfo{author}{\bibfnamefont{N.}~\bibnamefont{{Jarosik}}},
  \bibnamefont{et~al.}, \bibinfo{journal}{\apjs}
  \textbf{\bibinfo{volume}{180}}, \bibinfo{pages}{225} (\bibinfo{year}{2009}).

\bibitem[{\citenamefont{{Kowalski} et~al.}(2008)\citenamefont{{Kowalski},
  {Rubin}, {Aldering}, {Agostinho}, {Amadon}, {Amanullah}, {Balland},
  {Barbary}, {Blanc}, {Challis} et~al.}}]{Kowalski_2008}
\bibinfo{author}{\bibfnamefont{M.}~\bibnamefont{{Kowalski}}},
  \bibinfo{author}{\bibfnamefont{D.}~\bibnamefont{{Rubin}}},
  \bibinfo{author}{\bibfnamefont{G.}~\bibnamefont{{Aldering}}},
  \bibinfo{author}{\bibfnamefont{R.~J.} \bibnamefont{{Agostinho}}},
  \bibinfo{author}{\bibfnamefont{A.}~\bibnamefont{{Amadon}}},
  \bibinfo{author}{\bibfnamefont{R.}~\bibnamefont{{Amanullah}}},
  \bibinfo{author}{\bibfnamefont{C.}~\bibnamefont{{Balland}}},
  \bibinfo{author}{\bibfnamefont{K.}~\bibnamefont{{Barbary}}},
  \bibinfo{author}{\bibfnamefont{G.}~\bibnamefont{{Blanc}}},
  \bibinfo{author}{\bibfnamefont{P.~J.} \bibnamefont{{Challis}}},
  \bibnamefont{et~al.}, \bibinfo{journal}{\apj} \textbf{\bibinfo{volume}{686}},
  \bibinfo{pages}{749} (\bibinfo{year}{2008}).

\bibitem[{\citenamefont{{Daly} et~al.}(2009)\citenamefont{{Daly}, {Mory},
  {O'Dea}, {Kharb}, {Baum}, {Guerra}, and {Djorgovski}}}]{Daly_2009}
\bibinfo{author}{\bibfnamefont{R.~A.} \bibnamefont{{Daly}}},
  \bibinfo{author}{\bibfnamefont{M.~P.} \bibnamefont{{Mory}}},
  \bibinfo{author}{\bibfnamefont{C.~P.} \bibnamefont{{O'Dea}}},
  \bibinfo{author}{\bibfnamefont{P.}~\bibnamefont{{Kharb}}},
  \bibinfo{author}{\bibfnamefont{S.}~\bibnamefont{{Baum}}},
  \bibinfo{author}{\bibfnamefont{E.~J.} \bibnamefont{{Guerra}}},
  \bibnamefont{and} \bibinfo{author}{\bibfnamefont{S.~G.}
  \bibnamefont{{Djorgovski}}}, \bibinfo{journal}{\apj}
  \textbf{\bibinfo{volume}{691}}, \bibinfo{pages}{1058} (\bibinfo{year}{2009}).

\bibitem[{\citenamefont{{Clowe} et~al.}(2006)\citenamefont{{Clowe}, {Brada{\v
  c}}, {Gonzalez}, {Markevitch}, {Randall}, {Jones}, and
  {Zaritsky}}}]{Clowe_2006}
\bibinfo{author}{\bibfnamefont{D.}~\bibnamefont{{Clowe}}},
  \bibinfo{author}{\bibfnamefont{M.}~\bibnamefont{{Brada{\v c}}}},
  \bibinfo{author}{\bibfnamefont{A.~H.} \bibnamefont{{Gonzalez}}},
  \bibinfo{author}{\bibfnamefont{M.}~\bibnamefont{{Markevitch}}},
  \bibinfo{author}{\bibfnamefont{S.~W.} \bibnamefont{{Randall}}},
  \bibinfo{author}{\bibfnamefont{C.}~\bibnamefont{{Jones}}}, \bibnamefont{and}
  \bibinfo{author}{\bibfnamefont{D.}~\bibnamefont{{Zaritsky}}},
  \bibinfo{journal}{\apjl} \textbf{\bibinfo{volume}{648}},
  \bibinfo{pages}{L109} (\bibinfo{year}{2006}).

\bibitem[{\citenamefont{{Brada{\v c}} et~al.}(2008)\citenamefont{{Brada{\v c}},
  {Allen}, {Treu}, {Ebeling}, {Massey}, {Morris}, {von der Linden}, and
  {Applegate}}}]{Bradac_2008}
\bibinfo{author}{\bibfnamefont{M.}~\bibnamefont{{Brada{\v c}}}},
  \bibinfo{author}{\bibfnamefont{S.~W.} \bibnamefont{{Allen}}},
  \bibinfo{author}{\bibfnamefont{T.}~\bibnamefont{{Treu}}},
  \bibinfo{author}{\bibfnamefont{H.}~\bibnamefont{{Ebeling}}},
  \bibinfo{author}{\bibfnamefont{R.}~\bibnamefont{{Massey}}},
  \bibinfo{author}{\bibfnamefont{R.~G.} \bibnamefont{{Morris}}},
  \bibinfo{author}{\bibfnamefont{A.}~\bibnamefont{{von der Linden}}},
  \bibnamefont{and}
  \bibinfo{author}{\bibfnamefont{D.}~\bibnamefont{{Applegate}}},
  \bibinfo{journal}{\apj} \textbf{\bibinfo{volume}{687}}, \bibinfo{pages}{959}
  (\bibinfo{year}{2008}).

\bibitem[{\citenamefont{{Baltz}}(2004)}]{Baltz_2004}
\bibinfo{author}{\bibfnamefont{E.~A.} \bibnamefont{{Baltz}}},
  \bibinfo{journal}{arXiv:astro-ph/0412170}  (\bibinfo{year}{2004}).

\bibitem[{\citenamefont{{Bergstr{\"o}m}}(2009)}]{Bergstroem_2009_A}
\bibinfo{author}{\bibfnamefont{L.}~\bibnamefont{{Bergstr{\"o}m}}},
  \bibinfo{journal}{New Journal of Physics} \textbf{\bibinfo{volume}{11}},
  \bibinfo{pages}{105006} (\bibinfo{year}{2009}).

\bibitem[{\citenamefont{{Bertone} et~al.}(2005)\citenamefont{{Bertone},
  {Hooper}, and {Silk}}}]{Bertone_2005}
\bibinfo{author}{\bibfnamefont{G.}~\bibnamefont{{Bertone}}},
  \bibinfo{author}{\bibfnamefont{D.}~\bibnamefont{{Hooper}}}, \bibnamefont{and}
  \bibinfo{author}{\bibfnamefont{J.}~\bibnamefont{{Silk}}},
  \bibinfo{journal}{\physrep} \textbf{\bibinfo{volume}{405}},
  \bibinfo{pages}{279} (\bibinfo{year}{2005}).

\bibitem[{\citenamefont{{Taoso} et~al.}(2008)\citenamefont{{Taoso}, {Bertone},
  and {Masiero}}}]{Taoso_2008}
\bibinfo{author}{\bibfnamefont{M.}~\bibnamefont{{Taoso}}},
  \bibinfo{author}{\bibfnamefont{G.}~\bibnamefont{{Bertone}}},
  \bibnamefont{and}
  \bibinfo{author}{\bibfnamefont{A.}~\bibnamefont{{Masiero}}},
  \bibinfo{journal}{\jcap} \textbf{\bibinfo{volume}{3}}, \bibinfo{pages}{22}
  (\bibinfo{year}{2008}).

\bibitem[{\citenamefont{{Navarro} et~al.}(1996)\citenamefont{{Navarro},
  {Frenk}, and {White}}}]{Navarro_1996}
\bibinfo{author}{\bibfnamefont{J.~F.} \bibnamefont{{Navarro}}},
  \bibinfo{author}{\bibfnamefont{C.~S.} \bibnamefont{{Frenk}}},
  \bibnamefont{and} \bibinfo{author}{\bibfnamefont{S.~D.~M.}
  \bibnamefont{{White}}}, \bibinfo{journal}{\apj}
  \textbf{\bibinfo{volume}{462}}, \bibinfo{pages}{563} (\bibinfo{year}{1996}).

\bibitem[{\citenamefont{{Springel} et~al.}(2008)\citenamefont{{Springel},
  {Wang}, {Vogelsberger}, {Ludlow}, {Jenkins}, {Helmi}, {Navarro}, {Frenk}, and
  {White}}}]{Springel_2008}
\bibinfo{author}{\bibfnamefont{V.}~\bibnamefont{{Springel}}},
  \bibinfo{author}{\bibfnamefont{J.}~\bibnamefont{{Wang}}},
  \bibinfo{author}{\bibfnamefont{M.}~\bibnamefont{{Vogelsberger}}},
  \bibinfo{author}{\bibfnamefont{A.}~\bibnamefont{{Ludlow}}},
  \bibinfo{author}{\bibfnamefont{A.}~\bibnamefont{{Jenkins}}},
  \bibinfo{author}{\bibfnamefont{A.}~\bibnamefont{{Helmi}}},
  \bibinfo{author}{\bibfnamefont{J.~F.} \bibnamefont{{Navarro}}},
  \bibinfo{author}{\bibfnamefont{C.~S.} \bibnamefont{{Frenk}}},
  \bibnamefont{and} \bibinfo{author}{\bibfnamefont{S.~D.~M.}
  \bibnamefont{{White}}}, \bibinfo{journal}{\mnras}
  \textbf{\bibinfo{volume}{391}}, \bibinfo{pages}{1685} (\bibinfo{year}{2008}).

\bibitem[{\citenamefont{{Navarro} et~al.}(2010)\citenamefont{{Navarro},
  {Ludlow}, {Springel}, {Wang}, {Vogelsberger}, {White}, {Jenkins}, {Frenk},
  and {Helmi}}}]{Navarro_2010}
\bibinfo{author}{\bibfnamefont{J.~F.} \bibnamefont{{Navarro}}},
  \bibinfo{author}{\bibfnamefont{A.}~\bibnamefont{{Ludlow}}},
  \bibinfo{author}{\bibfnamefont{V.}~\bibnamefont{{Springel}}},
  \bibinfo{author}{\bibfnamefont{J.}~\bibnamefont{{Wang}}},
  \bibinfo{author}{\bibfnamefont{M.}~\bibnamefont{{Vogelsberger}}},
  \bibinfo{author}{\bibfnamefont{S.~D.~M.} \bibnamefont{{White}}},
  \bibinfo{author}{\bibfnamefont{A.}~\bibnamefont{{Jenkins}}},
  \bibinfo{author}{\bibfnamefont{C.~S.} \bibnamefont{{Frenk}}},
  \bibnamefont{and} \bibinfo{author}{\bibfnamefont{A.}~\bibnamefont{{Helmi}}},
  \bibinfo{journal}{\mnras} \textbf{\bibinfo{volume}{402}}, \bibinfo{pages}{21}
  (\bibinfo{year}{2010}).

\bibitem[{\citenamefont{{Klypin} et~al.}(1999)\citenamefont{{Klypin},
  {Kravtsov}, {Valenzuela}, and {Prada}}}]{Klypin_1999}
\bibinfo{author}{\bibfnamefont{A.}~\bibnamefont{{Klypin}}},
  \bibinfo{author}{\bibfnamefont{A.~V.} \bibnamefont{{Kravtsov}}},
  \bibinfo{author}{\bibfnamefont{O.}~\bibnamefont{{Valenzuela}}},
  \bibnamefont{and} \bibinfo{author}{\bibfnamefont{F.}~\bibnamefont{{Prada}}},
  \bibinfo{journal}{\apj} \textbf{\bibinfo{volume}{522}}, \bibinfo{pages}{82}
  (\bibinfo{year}{1999}).

\bibitem[{\citenamefont{{Moore} et~al.}(1999)\citenamefont{{Moore}, {Ghigna},
  {Governato}, {Lake}, {Quinn}, {Stadel}, and {Tozzi}}}]{Moore_1999}
\bibinfo{author}{\bibfnamefont{B.}~\bibnamefont{{Moore}}},
  \bibinfo{author}{\bibfnamefont{S.}~\bibnamefont{{Ghigna}}},
  \bibinfo{author}{\bibfnamefont{F.}~\bibnamefont{{Governato}}},
  \bibinfo{author}{\bibfnamefont{G.}~\bibnamefont{{Lake}}},
  \bibinfo{author}{\bibfnamefont{T.}~\bibnamefont{{Quinn}}},
  \bibinfo{author}{\bibfnamefont{J.}~\bibnamefont{{Stadel}}}, \bibnamefont{and}
  \bibinfo{author}{\bibfnamefont{P.}~\bibnamefont{{Tozzi}}},
  \bibinfo{journal}{\apjl} \textbf{\bibinfo{volume}{524}}, \bibinfo{pages}{L19}
  (\bibinfo{year}{1999}).

\bibitem[{\citenamefont{{Shapiro} et~al.}(2004)\citenamefont{{Shapiro},
  {Iliev}, and {Raga}}}]{Shapiro_2004}
\bibinfo{author}{\bibfnamefont{P.~R.} \bibnamefont{{Shapiro}}},
  \bibinfo{author}{\bibfnamefont{I.~T.} \bibnamefont{{Iliev}}},
  \bibnamefont{and} \bibinfo{author}{\bibfnamefont{A.~C.}
  \bibnamefont{{Raga}}}, \bibinfo{journal}{\mnras}
  \textbf{\bibinfo{volume}{348}}, \bibinfo{pages}{753} (\bibinfo{year}{2004}).

\bibitem[{\citenamefont{{Gilmore} et~al.}(2007)\citenamefont{{Gilmore},
  {Wilkinson}, {Wyse}, {Kleyna}, {Koch}, {Evans}, and {Grebel}}}]{Gilmore_2007}
\bibinfo{author}{\bibfnamefont{G.}~\bibnamefont{{Gilmore}}},
  \bibinfo{author}{\bibfnamefont{M.~I.} \bibnamefont{{Wilkinson}}},
  \bibinfo{author}{\bibfnamefont{R.~F.~G.} \bibnamefont{{Wyse}}},
  \bibinfo{author}{\bibfnamefont{J.~T.} \bibnamefont{{Kleyna}}},
  \bibinfo{author}{\bibfnamefont{A.}~\bibnamefont{{Koch}}},
  \bibinfo{author}{\bibfnamefont{N.~W.} \bibnamefont{{Evans}}},
  \bibnamefont{and} \bibinfo{author}{\bibfnamefont{E.~K.}
  \bibnamefont{{Grebel}}}, \bibinfo{journal}{\apj}
  \textbf{\bibinfo{volume}{663}}, \bibinfo{pages}{948} (\bibinfo{year}{2007}).

\bibitem[{\citenamefont{{Gentile} et~al.}(2004)\citenamefont{{Gentile},
  {Salucci}, {Klein}, {Vergani}, and {Kalberla}}}]{Gentile_2004}
\bibinfo{author}{\bibfnamefont{G.}~\bibnamefont{{Gentile}}},
  \bibinfo{author}{\bibfnamefont{P.}~\bibnamefont{{Salucci}}},
  \bibinfo{author}{\bibfnamefont{U.}~\bibnamefont{{Klein}}},
  \bibinfo{author}{\bibfnamefont{D.}~\bibnamefont{{Vergani}}},
  \bibnamefont{and}
  \bibinfo{author}{\bibfnamefont{P.}~\bibnamefont{{Kalberla}}},
  \bibinfo{journal}{\mnras} \textbf{\bibinfo{volume}{351}},
  \bibinfo{pages}{903} (\bibinfo{year}{2004}).

\bibitem[{\citenamefont{{Salucci} et~al.}(2007)\citenamefont{{Salucci}, {Lapi},
  {Tonini}, {Gentile}, {Yegorova}, and {Klein}}}]{Salucci_2007}
\bibinfo{author}{\bibfnamefont{P.}~\bibnamefont{{Salucci}}},
  \bibinfo{author}{\bibfnamefont{A.}~\bibnamefont{{Lapi}}},
  \bibinfo{author}{\bibfnamefont{C.}~\bibnamefont{{Tonini}}},
  \bibinfo{author}{\bibfnamefont{G.}~\bibnamefont{{Gentile}}},
  \bibinfo{author}{\bibfnamefont{I.}~\bibnamefont{{Yegorova}}},
  \bibnamefont{and} \bibinfo{author}{\bibfnamefont{U.}~\bibnamefont{{Klein}}},
  \bibinfo{journal}{\mnras} \textbf{\bibinfo{volume}{378}}, \bibinfo{pages}{41}
  (\bibinfo{year}{2007}).

\bibitem[{\citenamefont{{de Blok}}(2010)}]{deBlok_2010}
\bibinfo{author}{\bibfnamefont{W.~J.~G.} \bibnamefont{{de Blok}}},
  \bibinfo{journal}{Advances in Astronomy} \textbf{\bibinfo{volume}{2010}}
  (\bibinfo{year}{2010}).

\bibitem[{\citenamefont{{Spergel} and {Steinhardt}}(2000)}]{Spergel_2000}
\bibinfo{author}{\bibfnamefont{D.~N.} \bibnamefont{{Spergel}}}
  \bibnamefont{and} \bibinfo{author}{\bibfnamefont{P.~J.}
  \bibnamefont{{Steinhardt}}}, \bibinfo{journal}{\prl}
  \textbf{\bibinfo{volume}{84}}, \bibinfo{pages}{3760} (\bibinfo{year}{2000}).

\bibitem[{\citenamefont{{Yoshida} et~al.}(2000)\citenamefont{{Yoshida},
  {Springel}, {White}, and {Tormen}}}]{Yoshida_2000}
\bibinfo{author}{\bibfnamefont{N.}~\bibnamefont{{Yoshida}}},
  \bibinfo{author}{\bibfnamefont{V.}~\bibnamefont{{Springel}}},
  \bibinfo{author}{\bibfnamefont{S.~D.~M.} \bibnamefont{{White}}},
  \bibnamefont{and} \bibinfo{author}{\bibfnamefont{G.}~\bibnamefont{{Tormen}}},
  \bibinfo{journal}{\apjl} \textbf{\bibinfo{volume}{544}}, \bibinfo{pages}{L87}
  (\bibinfo{year}{2000}).

\bibitem[{\citenamefont{{D'Onghia} and {Burkert}}(2003)}]{Donghia_2003}
\bibinfo{author}{\bibfnamefont{E.}~\bibnamefont{{D'Onghia}}} \bibnamefont{and}
  \bibinfo{author}{\bibfnamefont{A.}~\bibnamefont{{Burkert}}},
  \bibinfo{journal}{\apj} \textbf{\bibinfo{volume}{586}}, \bibinfo{pages}{12}
  (\bibinfo{year}{2003}).

\bibitem[{\citenamefont{{Markevitch} et~al.}(2004)\citenamefont{{Markevitch},
  {Gonzalez}, {Clowe}, {Vikhlinin}, {Forman}, {Jones}, {Murray}, and
  {Tucker}}}]{Markevitch_2004}
\bibinfo{author}{\bibfnamefont{M.}~\bibnamefont{{Markevitch}}},
  \bibinfo{author}{\bibfnamefont{A.~H.} \bibnamefont{{Gonzalez}}},
  \bibinfo{author}{\bibfnamefont{D.}~\bibnamefont{{Clowe}}},
  \bibinfo{author}{\bibfnamefont{A.}~\bibnamefont{{Vikhlinin}}},
  \bibinfo{author}{\bibfnamefont{W.}~\bibnamefont{{Forman}}},
  \bibinfo{author}{\bibfnamefont{C.}~\bibnamefont{{Jones}}},
  \bibinfo{author}{\bibfnamefont{S.}~\bibnamefont{{Murray}}}, \bibnamefont{and}
  \bibinfo{author}{\bibfnamefont{W.}~\bibnamefont{{Tucker}}},
  \bibinfo{journal}{\apj} \textbf{\bibinfo{volume}{606}}, \bibinfo{pages}{819}
  (\bibinfo{year}{2004}).

\bibitem[{\citenamefont{{Randall} et~al.}(2008)\citenamefont{{Randall},
  {Markevitch}, {Clowe}, {Gonzalez}, and {Brada{\v c}}}}]{Randall_2008}
\bibinfo{author}{\bibfnamefont{S.~W.} \bibnamefont{{Randall}}},
  \bibinfo{author}{\bibfnamefont{M.}~\bibnamefont{{Markevitch}}},
  \bibinfo{author}{\bibfnamefont{D.}~\bibnamefont{{Clowe}}},
  \bibinfo{author}{\bibfnamefont{A.~H.} \bibnamefont{{Gonzalez}}},
  \bibnamefont{and} \bibinfo{author}{\bibfnamefont{M.}~\bibnamefont{{Brada{\v
  c}}}}, \bibinfo{journal}{\apj} \textbf{\bibinfo{volume}{679}},
  \bibinfo{pages}{1173} (\bibinfo{year}{2008}).

\bibitem[{\citenamefont{{Miralda-Escud{\'e}}}(2002)}]{Miralda_2002}
\bibinfo{author}{\bibfnamefont{J.}~\bibnamefont{{Miralda-Escud{\'e}}}},
  \bibinfo{journal}{\apj} \textbf{\bibinfo{volume}{564}}, \bibinfo{pages}{60}
  (\bibinfo{year}{2002}).

\bibitem[{\citenamefont{{Bode} et~al.}(2001)\citenamefont{{Bode}, {Ostriker},
  and {Turok}}}]{Bode_2001}
\bibinfo{author}{\bibfnamefont{P.}~\bibnamefont{{Bode}}},
  \bibinfo{author}{\bibfnamefont{J.~P.} \bibnamefont{{Ostriker}}},
  \bibnamefont{and} \bibinfo{author}{\bibfnamefont{N.}~\bibnamefont{{Turok}}},
  \bibinfo{journal}{\apj} \textbf{\bibinfo{volume}{556}}, \bibinfo{pages}{93}
  (\bibinfo{year}{2001}).

\bibitem[{\citenamefont{{Sommer-Larsen} and {Dolgov}}(2001)}]{Sommer_2001}
\bibinfo{author}{\bibfnamefont{J.}~\bibnamefont{{Sommer-Larsen}}}
  \bibnamefont{and} \bibinfo{author}{\bibfnamefont{A.}~\bibnamefont{{Dolgov}}},
  \bibinfo{journal}{\apj} \textbf{\bibinfo{volume}{551}}, \bibinfo{pages}{608}
  (\bibinfo{year}{2001}).

\bibitem[{\citenamefont{{Viel} et~al.}(2008)\citenamefont{{Viel}, {Becker},
  {Bolton}, {Haehnelt}, {Rauch}, and {Sargent}}}]{Viel_2008}
\bibinfo{author}{\bibfnamefont{M.}~\bibnamefont{{Viel}}},
  \bibinfo{author}{\bibfnamefont{G.~D.} \bibnamefont{{Becker}}},
  \bibinfo{author}{\bibfnamefont{J.~S.} \bibnamefont{{Bolton}}},
  \bibinfo{author}{\bibfnamefont{M.~G.} \bibnamefont{{Haehnelt}}},
  \bibinfo{author}{\bibfnamefont{M.}~\bibnamefont{{Rauch}}}, \bibnamefont{and}
  \bibinfo{author}{\bibfnamefont{W.~L.~W.} \bibnamefont{{Sargent}}},
  \bibinfo{journal}{\prl} \textbf{\bibinfo{volume}{100}},
  \bibinfo{pages}{041304} (\bibinfo{year}{2008}).

\bibitem[{\citenamefont{{Miranda} and {Macci{\`o}}}(2007)}]{Miranda_2007}
\bibinfo{author}{\bibfnamefont{M.}~\bibnamefont{{Miranda}}} \bibnamefont{and}
  \bibinfo{author}{\bibfnamefont{A.~V.} \bibnamefont{{Macci{\`o}}}},
  \bibinfo{journal}{\mnras} \textbf{\bibinfo{volume}{382}},
  \bibinfo{pages}{1225} (\bibinfo{year}{2007}).

\bibitem[{\citenamefont{{Boyarsky} et~al.}(2009)\citenamefont{{Boyarsky},
  {Lesgourgues}, {Ruchayskiy}, and {Viel}}}]{Boyarsky_2009}
\bibinfo{author}{\bibfnamefont{A.}~\bibnamefont{{Boyarsky}}},
  \bibinfo{author}{\bibfnamefont{J.}~\bibnamefont{{Lesgourgues}}},
  \bibinfo{author}{\bibfnamefont{O.}~\bibnamefont{{Ruchayskiy}}},
  \bibnamefont{and} \bibinfo{author}{\bibfnamefont{M.}~\bibnamefont{{Viel}}},
  \bibinfo{journal}{\jcap} \textbf{\bibinfo{volume}{5}}, \bibinfo{pages}{12}
  (\bibinfo{year}{2009}).

\bibitem[{\citenamefont{{Boyanovsky} et~al.}(2008)\citenamefont{{Boyanovsky},
  {de Vega}, and {Sanchez}}}]{Boyanovsky_2008}
\bibinfo{author}{\bibfnamefont{D.}~\bibnamefont{{Boyanovsky}}},
  \bibinfo{author}{\bibfnamefont{H.~J.} \bibnamefont{{de Vega}}},
  \bibnamefont{and} \bibinfo{author}{\bibfnamefont{N.~G.}
  \bibnamefont{{Sanchez}}}, \bibinfo{journal}{\prd}
  \textbf{\bibinfo{volume}{77}}, \bibinfo{pages}{043518}
  (\bibinfo{year}{2008}).

\bibitem[{\citenamefont{{Kaplinghat} et~al.}(2000)\citenamefont{{Kaplinghat},
  {Knox}, and {Turner}}}]{Kaplinghat_2000}
\bibinfo{author}{\bibfnamefont{M.}~\bibnamefont{{Kaplinghat}}},
  \bibinfo{author}{\bibfnamefont{L.}~\bibnamefont{{Knox}}}, \bibnamefont{and}
  \bibinfo{author}{\bibfnamefont{M.~S.} \bibnamefont{{Turner}}},
  \bibinfo{journal}{\prl} \textbf{\bibinfo{volume}{85}}, \bibinfo{pages}{3335}
  (\bibinfo{year}{2000}).

\bibitem[{\citenamefont{{Beacom} et~al.}(2007)\citenamefont{{Beacom}, {Bell},
  and {Mack}}}]{Beacom_2007}
\bibinfo{author}{\bibfnamefont{J.~F.} \bibnamefont{{Beacom}}},
  \bibinfo{author}{\bibfnamefont{N.~F.} \bibnamefont{{Bell}}},
  \bibnamefont{and} \bibinfo{author}{\bibfnamefont{G.~D.}
  \bibnamefont{{Mack}}}, \bibinfo{journal}{\prl} \textbf{\bibinfo{volume}{99}},
  \bibinfo{pages}{231301} (\bibinfo{year}{2007}).

\bibitem[{\citenamefont{{Baldi} et~al.}(2010)\citenamefont{{Baldi},
  {Pettorino}, {Robbers}, and {Springel}}}]{Baldi_2010}
\bibinfo{author}{\bibfnamefont{M.}~\bibnamefont{{Baldi}}},
  \bibinfo{author}{\bibfnamefont{V.}~\bibnamefont{{Pettorino}}},
  \bibinfo{author}{\bibfnamefont{G.}~\bibnamefont{{Robbers}}},
  \bibnamefont{and}
  \bibinfo{author}{\bibfnamefont{V.}~\bibnamefont{{Springel}}},
  \bibinfo{journal}{\mnras} \textbf{\bibinfo{volume}{403}},
  \bibinfo{pages}{1684} (\bibinfo{year}{2010}).

\bibitem[{\citenamefont{{Cyburt} et~al.}(2002)\citenamefont{{Cyburt}, {Fields},
  {Pavlidou}, and {Wandelt}}}]{Cyburt_2002}
\bibinfo{author}{\bibfnamefont{R.~H.} \bibnamefont{{Cyburt}}},
  \bibinfo{author}{\bibfnamefont{B.~D.} \bibnamefont{{Fields}}},
  \bibinfo{author}{\bibfnamefont{V.}~\bibnamefont{{Pavlidou}}},
  \bibnamefont{and}
  \bibinfo{author}{\bibfnamefont{B.}~\bibnamefont{{Wandelt}}},
  \bibinfo{journal}{\prd} \textbf{\bibinfo{volume}{65}},
  \bibinfo{pages}{123503} (\bibinfo{year}{2002}).

\bibitem[{\citenamefont{{Ackerman} et~al.}(2009)\citenamefont{{Ackerman},
  {Buckley}, {Carroll}, and {Kamionkowski}}}]{Ackerman_2009}
\bibinfo{author}{\bibfnamefont{L.}~\bibnamefont{{Ackerman}}},
  \bibinfo{author}{\bibfnamefont{M.~R.} \bibnamefont{{Buckley}}},
  \bibinfo{author}{\bibfnamefont{S.~M.} \bibnamefont{{Carroll}}},
  \bibnamefont{and}
  \bibinfo{author}{\bibfnamefont{M.}~\bibnamefont{{Kamionkowski}}},
  \bibinfo{journal}{\prd} \textbf{\bibinfo{volume}{79}},
  \bibinfo{pages}{023519} (\bibinfo{year}{2009}).

\bibitem[{\citenamefont{{Grasso} et~al.}(2009)\citenamefont{{Grasso},
  {Profumo}, {Strong}, {Baldini}, {Bellazzini}, {Bloom}, {Bregeon}, {di
  Bernardo}, {Gaggero}, {Giglietto} et~al.}}]{Grasso_2009}
\bibinfo{author}{\bibfnamefont{D.}~\bibnamefont{{Grasso}}},
  \bibinfo{author}{\bibfnamefont{S.}~\bibnamefont{{Profumo}}},
  \bibinfo{author}{\bibfnamefont{A.~W.} \bibnamefont{{Strong}}},
  \bibinfo{author}{\bibfnamefont{L.}~\bibnamefont{{Baldini}}},
  \bibinfo{author}{\bibfnamefont{R.}~\bibnamefont{{Bellazzini}}},
  \bibinfo{author}{\bibfnamefont{E.~D.} \bibnamefont{{Bloom}}},
  \bibinfo{author}{\bibfnamefont{J.}~\bibnamefont{{Bregeon}}},
  \bibinfo{author}{\bibfnamefont{G.}~\bibnamefont{{di Bernardo}}},
  \bibinfo{author}{\bibfnamefont{D.}~\bibnamefont{{Gaggero}}},
  \bibinfo{author}{\bibfnamefont{N.}~\bibnamefont{{Giglietto}}},
  \bibnamefont{et~al.}, \bibinfo{journal}{Astroparticle Physics}
  \textbf{\bibinfo{volume}{32}}, \bibinfo{pages}{140} (\bibinfo{year}{2009}).

\bibitem[{\citenamefont{{Bergstr{\"o}m}
  et~al.}(2009)\citenamefont{{Bergstr{\"o}m}, {Edsj{\"o}}, and
  {Zaharijas}}}]{Bergstroem_2009_B}
\bibinfo{author}{\bibfnamefont{L.}~\bibnamefont{{Bergstr{\"o}m}}},
  \bibinfo{author}{\bibfnamefont{J.}~\bibnamefont{{Edsj{\"o}}}},
  \bibnamefont{and}
  \bibinfo{author}{\bibfnamefont{G.}~\bibnamefont{{Zaharijas}}},
  \bibinfo{journal}{\prl} \textbf{\bibinfo{volume}{103}},
  \bibinfo{pages}{031103} (\bibinfo{year}{2009}).

\bibitem[{\citenamefont{{Narain} et~al.}(2006)\citenamefont{{Narain},
  {Schaffner-Bielich}, and {Mishustin}}}]{Narain_2006}
\bibinfo{author}{\bibfnamefont{G.}~\bibnamefont{{Narain}}},
  \bibinfo{author}{\bibfnamefont{J.}~\bibnamefont{{Schaffner-Bielich}}},
  \bibnamefont{and} \bibinfo{author}{\bibfnamefont{I.~N.}
  \bibnamefont{{Mishustin}}}, \bibinfo{journal}{\prd}
  \textbf{\bibinfo{volume}{74}}, \bibinfo{pages}{063003}
  (\bibinfo{year}{2006}).

\bibitem[{\citenamefont{{Agnihotri} et~al.}(2009)\citenamefont{{Agnihotri},
  {Schaffner-Bielich}, and {Mishustin}}}]{Agnihotri_2009}
\bibinfo{author}{\bibfnamefont{P.}~\bibnamefont{{Agnihotri}}},
  \bibinfo{author}{\bibfnamefont{J.}~\bibnamefont{{Schaffner-Bielich}}},
  \bibnamefont{and} \bibinfo{author}{\bibfnamefont{I.~N.}
  \bibnamefont{{Mishustin}}}, \bibinfo{journal}{\prd}
  \textbf{\bibinfo{volume}{79}}, \bibinfo{pages}{084033}
  (\bibinfo{year}{2009}).

\bibitem[{\citenamefont{{Boyle} and {Buonanno}}(2008)}]{Boyle_2008}
\bibinfo{author}{\bibfnamefont{L.~A.} \bibnamefont{{Boyle}}} \bibnamefont{and}
  \bibinfo{author}{\bibfnamefont{A.}~\bibnamefont{{Buonanno}}},
  \bibinfo{journal}{\prd} \textbf{\bibinfo{volume}{78}},
  \bibinfo{pages}{043531} (\bibinfo{year}{2008}).

\bibitem[{\citenamefont{{Boeckel} and
  {Schaffner-Bielich}}(2007)}]{Boeckel_2007}
\bibinfo{author}{\bibfnamefont{T.}~\bibnamefont{{Boeckel}}} \bibnamefont{and}
  \bibinfo{author}{\bibfnamefont{J.}~\bibnamefont{{Schaffner-Bielich}}},
  \bibinfo{journal}{\prd} \textbf{\bibinfo{volume}{76}},
  \bibinfo{pages}{103509} (\bibinfo{year}{2007}).

\bibitem[{\citenamefont{{Mather} et~al.}(1999)\citenamefont{{Mather}, {Fixsen},
  {Shafer}, {Mosier}, and {Wilkinson}}}]{Mather_1999}
\bibinfo{author}{\bibfnamefont{J.~C.} \bibnamefont{{Mather}}},
  \bibinfo{author}{\bibfnamefont{D.~J.} \bibnamefont{{Fixsen}}},
  \bibinfo{author}{\bibfnamefont{R.~A.} \bibnamefont{{Shafer}}},
  \bibinfo{author}{\bibfnamefont{C.}~\bibnamefont{{Mosier}}}, \bibnamefont{and}
  \bibinfo{author}{\bibfnamefont{D.~T.} \bibnamefont{{Wilkinson}}},
  \bibinfo{journal}{\apj} \textbf{\bibinfo{volume}{512}}, \bibinfo{pages}{511}
  (\bibinfo{year}{1999}).

\bibitem[{\citenamefont{{Brandenberger}
  et~al.}(2007)\citenamefont{{Brandenberger}, {Nayeri}, {Patil}, and
  {Vafa}}}]{Brandenberger_2007_1}
\bibinfo{author}{\bibfnamefont{R.~H.} \bibnamefont{{Brandenberger}}},
  \bibinfo{author}{\bibfnamefont{A.}~\bibnamefont{{Nayeri}}},
  \bibinfo{author}{\bibfnamefont{S.~P.} \bibnamefont{{Patil}}},
  \bibnamefont{and} \bibinfo{author}{\bibfnamefont{C.}~\bibnamefont{{Vafa}}},
  \bibinfo{journal}{\ijmpa} \textbf{\bibinfo{volume}{22}},
  \bibinfo{pages}{3621} (\bibinfo{year}{2007}).

\bibitem[{\citenamefont{{Brandenberger}}(2007)}]{Brandenberger_2007_2}
\bibinfo{author}{\bibfnamefont{R.~H.} \bibnamefont{{Brandenberger}}},
  \bibinfo{journal}{\ptps} \textbf{\bibinfo{volume}{171}}, \bibinfo{pages}{121}
  (\bibinfo{year}{2007}).

\bibitem[{\citenamefont{{Simha} and {Steigman}}(2008)}]{Simha_2008}
\bibinfo{author}{\bibfnamefont{V.}~\bibnamefont{{Simha}}} \bibnamefont{and}
  \bibinfo{author}{\bibfnamefont{G.}~\bibnamefont{{Steigman}}},
  \bibinfo{journal}{\jcap} \textbf{\bibinfo{volume}{6}}, \bibinfo{pages}{16}
  (\bibinfo{year}{2008}).

\bibitem[{\citenamefont{{Steigman}}(2007)}]{Steigman_2007}
\bibinfo{author}{\bibfnamefont{G.}~\bibnamefont{{Steigman}}},
  \bibinfo{journal}{Annual Review of Nuclear and Particle Science}
  \textbf{\bibinfo{volume}{57}}, \bibinfo{pages}{463} (\bibinfo{year}{2007}).

\bibitem[{\citenamefont{{Hogan} and {Dalcanton}}(2000)}]{Hogan_2000}
\bibinfo{author}{\bibfnamefont{C.~J.} \bibnamefont{{Hogan}}} \bibnamefont{and}
  \bibinfo{author}{\bibfnamefont{J.~J.} \bibnamefont{{Dalcanton}}},
  \bibinfo{journal}{\prd} \textbf{\bibinfo{volume}{62}},
  \bibinfo{pages}{063511} (\bibinfo{year}{2000}).

\bibitem[{\citenamefont{{Hui}}(2001)}]{Hui_2001}
\bibinfo{author}{\bibfnamefont{L.}~\bibnamefont{{Hui}}},
  \bibinfo{journal}{\prl} \textbf{\bibinfo{volume}{86}}, \bibinfo{pages}{3467}
  (\bibinfo{year}{2001}).

\bibitem[{\citenamefont{{Jungman} et~al.}(1996)\citenamefont{{Jungman},
  {Kamionkowski}, and {Griest}}}]{Jungman_1996}
\bibinfo{author}{\bibfnamefont{G.}~\bibnamefont{{Jungman}}},
  \bibinfo{author}{\bibfnamefont{M.}~\bibnamefont{{Kamionkowski}}},
  \bibnamefont{and} \bibinfo{author}{\bibfnamefont{K.}~\bibnamefont{{Griest}}},
  \bibinfo{journal}{\physrep} \textbf{\bibinfo{volume}{267}},
  \bibinfo{pages}{195} (\bibinfo{year}{1996}).

\bibitem[{\citenamefont{{Griest} and {Kamionkowski}}(1990)}]{Griest_1990}
\bibinfo{author}{\bibfnamefont{K.}~\bibnamefont{{Griest}}} \bibnamefont{and}
  \bibinfo{author}{\bibfnamefont{M.}~\bibnamefont{{Kamionkowski}}},
  \bibinfo{journal}{\prl} \textbf{\bibinfo{volume}{64}}, \bibinfo{pages}{615}
  (\bibinfo{year}{1990}).

\bibitem[{\citenamefont{{Y{\"u}ksel} et~al.}(2007)\citenamefont{{Y{\"u}ksel},
  {Horiuchi}, {Beacom}, and {Ando}}}]{Yueksel_2007}
\bibinfo{author}{\bibfnamefont{H.}~\bibnamefont{{Y{\"u}ksel}}},
  \bibinfo{author}{\bibfnamefont{S.}~\bibnamefont{{Horiuchi}}},
  \bibinfo{author}{\bibfnamefont{J.~F.} \bibnamefont{{Beacom}}},
  \bibnamefont{and} \bibinfo{author}{\bibfnamefont{S.}~\bibnamefont{{Ando}}},
  \bibinfo{journal}{\prd} \textbf{\bibinfo{volume}{76}},
  \bibinfo{pages}{123506} (\bibinfo{year}{2007}).

\bibitem[{\citenamefont{{Catena} et~al.}(2009)\citenamefont{{Catena},
  {Fornengo}, {Pato}, {Pieri}, and {Masiero}}}]{Catena_2009}
\bibinfo{author}{\bibfnamefont{R.}~\bibnamefont{{Catena}}},
  \bibinfo{author}{\bibfnamefont{N.}~\bibnamefont{{Fornengo}}},
  \bibinfo{author}{\bibfnamefont{M.}~\bibnamefont{{Pato}}},
  \bibinfo{author}{\bibfnamefont{L.}~\bibnamefont{{Pieri}}}, \bibnamefont{and}
  \bibinfo{author}{\bibfnamefont{A.}~\bibnamefont{{Masiero}}},
  \bibinfo{journal}{arXiv:0912.4421~[astro-ph.CO]}  (\bibinfo{year}{2009}).

\bibitem[{\citenamefont{{Schmid} et~al.}(1999)\citenamefont{{Schmid},
  {Schwarz}, and {Widerin}}}]{Schmid_1999}
\bibinfo{author}{\bibfnamefont{C.}~\bibnamefont{{Schmid}}},
  \bibinfo{author}{\bibfnamefont{D.~J.} \bibnamefont{{Schwarz}}},
  \bibnamefont{and}
  \bibinfo{author}{\bibfnamefont{P.}~\bibnamefont{{Widerin}}},
  \bibinfo{journal}{\prd} \textbf{\bibinfo{volume}{59}},
  \bibinfo{pages}{043517} (\bibinfo{year}{1999}).

\bibitem[{\citenamefont{{Hwang}}(1993)}]{Hwang_1993}
\bibinfo{author}{\bibfnamefont{J.}~\bibnamefont{{Hwang}}},
  \bibinfo{journal}{\apj} \textbf{\bibinfo{volume}{415}}, \bibinfo{pages}{486}
  (\bibinfo{year}{1993}).

\bibitem[{\citenamefont{{Boehm} and {Schaeffer}}(2005)}]{Boehm_2005}
\bibinfo{author}{\bibfnamefont{C.}~\bibnamefont{{Boehm}}} \bibnamefont{and}
  \bibinfo{author}{\bibfnamefont{R.}~\bibnamefont{{Schaeffer}}},
  \bibinfo{journal}{\aap} \textbf{\bibinfo{volume}{438}}, \bibinfo{pages}{419}
  (\bibinfo{year}{2005}).

\bibitem[{\citenamefont{{Boehm} et~al.}(2002)\citenamefont{{Boehm}, {Fayet},
  and {Schaeffer}}}]{Boehm_2002}
\bibinfo{author}{\bibfnamefont{C.}~\bibnamefont{{Boehm}}},
  \bibinfo{author}{\bibfnamefont{P.}~\bibnamefont{{Fayet}}}, \bibnamefont{and}
  \bibinfo{author}{\bibfnamefont{R.}~\bibnamefont{{Schaeffer}}}, in
  \emph{\bibinfo{booktitle}{Dark Matter in Astro- and Particle Physics, DARK
  2002}}, edited by \bibinfo{editor}{\bibnamefont{{H.~V.~Klapdor-Kleingrothaus
  \& R.~D.~Viollier}}} (\bibinfo{year}{2002}), pp. \bibinfo{pages}{333--344},
  \bibinfo{note}{arXiv:astro-ph/0205406}.

\bibitem[{\citenamefont{{Alcock} et~al.}(1998)\citenamefont{{Alcock},
  {Allsman}, {Alves}, {Ansari}, {Aubourg}, {Axelrod}, {Bareyre}, {Beaulieu},
  {Becker}, {Bennett} et~al.}}]{Alcock_1998}
\bibinfo{author}{\bibfnamefont{C.}~\bibnamefont{{Alcock}}},
  \bibinfo{author}{\bibfnamefont{R.~A.} \bibnamefont{{Allsman}}},
  \bibinfo{author}{\bibfnamefont{D.}~\bibnamefont{{Alves}}},
  \bibinfo{author}{\bibfnamefont{R.}~\bibnamefont{{Ansari}}},
  \bibinfo{author}{\bibfnamefont{E.}~\bibnamefont{{Aubourg}}},
  \bibinfo{author}{\bibfnamefont{T.~S.} \bibnamefont{{Axelrod}}},
  \bibinfo{author}{\bibfnamefont{P.}~\bibnamefont{{Bareyre}}},
  \bibinfo{author}{\bibfnamefont{J.}~\bibnamefont{{Beaulieu}}},
  \bibinfo{author}{\bibfnamefont{A.~C.} \bibnamefont{{Becker}}},
  \bibinfo{author}{\bibfnamefont{D.~P.} \bibnamefont{{Bennett}}},
  \bibnamefont{et~al.}, \bibinfo{journal}{\apjl}
  \textbf{\bibinfo{volume}{499}}, \bibinfo{pages}{L9+} (\bibinfo{year}{1998}).

\bibitem[{\citenamefont{{Afonso} et~al.}(2003)\citenamefont{{Afonso}, {Albert},
  {Andersen}, {Ansari}, {Aubourg}, {Bareyre}, {Beaulieu}, {Blanc}, {Charlot},
  {Couchot} et~al.}}]{Afonso_2003}
\bibinfo{author}{\bibfnamefont{C.}~\bibnamefont{{Afonso}}},
  \bibinfo{author}{\bibfnamefont{J.~N.} \bibnamefont{{Albert}}},
  \bibinfo{author}{\bibfnamefont{J.}~\bibnamefont{{Andersen}}},
  \bibinfo{author}{\bibfnamefont{R.}~\bibnamefont{{Ansari}}},
  \bibinfo{author}{\bibfnamefont{{\'E}.}~\bibnamefont{{Aubourg}}},
  \bibinfo{author}{\bibfnamefont{P.}~\bibnamefont{{Bareyre}}},
  \bibinfo{author}{\bibfnamefont{J.~P.} \bibnamefont{{Beaulieu}}},
  \bibinfo{author}{\bibfnamefont{G.}~\bibnamefont{{Blanc}}},
  \bibinfo{author}{\bibfnamefont{X.}~\bibnamefont{{Charlot}}},
  \bibinfo{author}{\bibfnamefont{F.}~\bibnamefont{{Couchot}}},
  \bibnamefont{et~al.}, \bibinfo{journal}{\aap} \textbf{\bibinfo{volume}{400}},
  \bibinfo{pages}{951} (\bibinfo{year}{2003}).

\bibitem[{\citenamefont{{Bringmann} and {Hofmann}}(2007)}]{Bringmann_2007}
\bibinfo{author}{\bibfnamefont{T.}~\bibnamefont{{Bringmann}}} \bibnamefont{and}
  \bibinfo{author}{\bibfnamefont{S.}~\bibnamefont{{Hofmann}}},
  \bibinfo{journal}{\jcap} \textbf{\bibinfo{volume}{4}}, \bibinfo{pages}{16}
  (\bibinfo{year}{2007}).

\bibitem[{\citenamefont{{Mukhanov}}(2004)}]{Mukhanov_2004}
\bibinfo{author}{\bibfnamefont{V.}~\bibnamefont{{Mukhanov}}},
  \bibinfo{journal}{International Journal of Theoretical Physics}
  \textbf{\bibinfo{volume}{43}}, \bibinfo{pages}{669} (\bibinfo{year}{2004}).

\bibitem[{\citenamefont{{Mukhanov}}(2005)}]{Mukhanov_2005}
\bibinfo{author}{\bibfnamefont{V.}~\bibnamefont{{Mukhanov}}},
  \emph{\bibinfo{title}{{Physical Foundations of Cosmology}}}
  (\bibinfo{year}{2005}).

\bibitem[{\citenamefont{{Aver} et~al.}(2010)\citenamefont{{Aver}, {Olive}, and
  {Skillman}}}]{Aver_2010}
\bibinfo{author}{\bibfnamefont{E.}~\bibnamefont{{Aver}}},
  \bibinfo{author}{\bibfnamefont{K.~A.} \bibnamefont{{Olive}}},
  \bibnamefont{and} \bibinfo{author}{\bibfnamefont{E.~D.}
  \bibnamefont{{Skillman}}}, \bibinfo{journal}{\jcap}
  \textbf{\bibinfo{volume}{5}}, \bibinfo{pages}{3} (\bibinfo{year}{2010}).

\bibitem[{\citenamefont{{Izotov} and {Thuan}}(2010)}]{Izotov_2010}
\bibinfo{author}{\bibfnamefont{Y.~I.} \bibnamefont{{Izotov}}} \bibnamefont{and}
  \bibinfo{author}{\bibfnamefont{T.~X.} \bibnamefont{{Thuan}}},
  \bibinfo{journal}{\apjl} \textbf{\bibinfo{volume}{710}}, \bibinfo{pages}{L67}
  (\bibinfo{year}{2010}).

\end{thebibliography}

\end{document}